\documentclass{ieeeaccess}
\usepackage{amsmath,amssymb,amsfonts}
\usepackage{algorithmic}
\usepackage{cite}
\usepackage{textcomp}
\usepackage{graphicx}
\usepackage{dcolumn}
\usepackage{bm}
\usepackage{comment}
\usepackage{mathtools}
\usepackage{resizegather}
\usepackage{balance}
\usepackage{tensor}
\usepackage{comment}
\usepackage{hyperref}

\newcommand{\ket}[1]{\left|#1\right\rangle}
\newcommand{\bra}[1]{\left\langle#1\right|}
\newcommand{\braket}[2]{\left\langle#1\middle|#2\right\rangle}

\begin{document}
\history{Date of publication xxxx 00, 0000, date of current version Nov. 21, 2019.}
\doi{10.1109/ACCESS.2019.2960684}

\title{CMOS Position-Based Charge Qubits: Theoretical Analysis of Control and Entanglement}

\author{
\uppercase{Elena Blokhina}\authorrefmark{1} \IEEEmembership{Senior Member, IEEE},
 \uppercase{Panagiotis Giounanlis}\authorrefmark{*,1} \IEEEmembership{Member, IEEE},
 \uppercase{Andrew Mitchell}\authorrefmark{3},
 \uppercase{Dirk R. Leipold}\authorrefmark{2} \IEEEmembership{Member, IEEE}, and
\uppercase{Robert Bogdan Staszewski}\authorrefmark{1,2}
\IEEEmembership{Fellow, IEEE}}
\address[1]{School of Electrical and Electronic Engineering, University College Dublin, Dublin 4, Ireland (e-mail: elena.blokhina@ucd.ie)}
\address[2]{Equal1 Labs, Fremont,
CA 94536, USA (e-mail: dirk.leipold@equal1.com)}
\address[3]{School of Physics, University College Dublin, Belfield, Dublin 4, Ireland (e-mail: andrew.mitchell@ucd.ie)}
\tfootnote{
This work was supported by Science Foundation Ireland under Grant 14/RP/I2921.\\
\textbf{IEEE Xplore url:}\url{https://ieeexplore.ieee.org/document/8936906}}
\markboth
{Blokhina \headeretal: CMOS Position-Based Charge Qubits: Theoretical Analysis of Control and Entanglement}
{Blokhina \headeretal: CMOS Charge Qubits: Theoretical Analysis of Control and Entanglement}

\corresp{$^*$Corresponding author: Panagiotis Giounanlis (e-mail: panagiotis.giounanlis@ucd.ie).}

\begin{abstract}
In this study, a formal definition, robustness analysis and discussion on the control of a position-based semiconductor charge qubit are presented. Such a qubit can be realized in a chain of coupled quantum dots, forming a register of charge-coupled transistor-like devices, and is intended for CMOS implementation in scalable quantum computers. We discuss the construction and operation of this qubit, its Bloch sphere, and relation with maximally localized Wannier functions which define its position-based nature. We then demonstrate how to build a tight-binding model of single and multiple interacting qubits from first principles of the Schr\"odinger formalism. We provide all required formulae to calculate the maximally localized functions and the entries of the Hamiltonian matrix in the presence of interaction between qubits. We use three illustrative examples to demonstrate the electrostatic interaction of electrons and discuss how to build a model for many-electron (qubit) system. To conclude this study, we show that charge qubits can be entangled through electrostatic interaction.

\end{abstract}

\begin{IEEEkeywords}
CMOS technology, charge qubit, position-based qubit, electrostatically controlled qubit, single-electron devices, tight-binding formalism, Schr\"odinger formalism, entanglement, entanglement entropy, Bloch sphere.
\end{IEEEkeywords}

\titlepgskip=-15pt

\maketitle

\section{Introduction}

Quantum computing is a still-emerging paradigm that utilizes the fundamental principles of quantum mechanics such as superposition and entanglement. The range of complex problems from mathematics, chemistry and material science that could be solved with quantum computing is immense~\cite{Reilly_2015,Van_2016,Vandersypen_2017}. A quantum bit (qubit) is the basic unit of quantum information, typically comprising a nanoscale quantum mechanical two-state system. The qubits are extremely fragile and difficult to manipulate and read out, since all the operations until the final read-out must be done non-destructively to preserve the crucial property of quantum coherence. They typically require extremely low, cryogenic temperatures to operate in order to preserve their coherent superposition state. Furthermore, quantum computation requires a fault-tolerant manipulation of many coupled qubits, making the search for robust qubits suitable for mass production a necessity. In this regard, an appealing paradigm for scalable quantum computation would exploit existing advanced semiconductor technologies.

The quantum computer differs from the classical digital computer in the sense that instead of using a binary digit (bit) to represent Boolean logic states of `0' or `1', it uses a qubit which can be in a superposition of quantum counterpart states of $\ket 0$ and $\ket 1$. Among a number of proposed technologies for realizing quantum computation, one can highlight the following ones. Possibly the most promising approach from the point of view of decoherence time is based on trapped ions~\cite{Debnath_2016}. It has been recently demonstrated that trapped ions and photons can achieve a large number of entangled qubits~\cite{Friis_2018}. Unfortunately, they are difficult to manipulate and are perceived as rather unsuited for a very large-scale integration (although the research into trapped ions is actively ongoing). Solid-state methods, on the other hand, rely on the usage of collective physical phenomena of macroscopic quantum states as, for instance, observed in superconductors or in superfluids~\cite{Nori_2005}. Integrated, on-chip solutions based on superconductors (employing Josephson junctions) are currently the dominant technology with the number of qubits ranging from 19 to 72 and operating at extremely low temperatures of 15~mK. There are many other promising theoretical proposals based on topological quantum computation by braiding Majorana fermions or other non-Abelian quasiparticles in condensed matter systems~\cite{nayak2008non}. However, none of these proposals have been experimentally realized to date. Some further discussion on state-of-the-art of qubits can be found in~\cite{Giounanlis_2019}.

\begingroup
\newlength{\xfigwd}
\setlength{\xfigwd}{\textwidth}
\endgroup

\begin{figure}[t!]
\centering
\includegraphics[width=0.5\columnwidth]{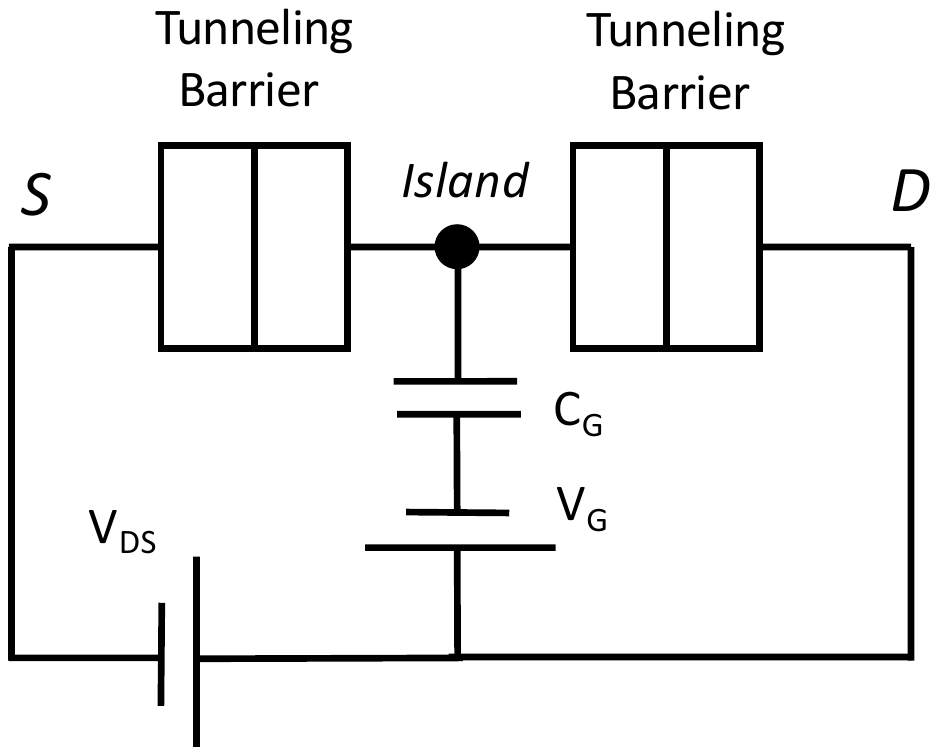}
\caption{Schematic diagram of a single-electron transistor (SET), a device allowing a single injected electron to be manipulated via the tunneling barriers by controlling voltages applied at the source (S), drain (D) and gate (G) terminals.}
\label{fig:SET}
\vspace{0mm}
\end{figure}

Over recent years, there has been ongoing research on semiconductor qubits~\cite{Fujisawa_2004,Fujisawa2,Petta_2010,Weichselbaum_2004,Szafran_2018, Ihara_2015,Giounanlis_2019, Leipold_2019, esscirc_2019} due to their promising compatibility with batch fabrication and enormous progress in CMOS fabrication technologies. Inspired by previous and very recent works on charge-based semiconductor qubits~\cite{Fujisawa2,Petta_2010,Hensgens_2017,Goldmann_2014,Giounanlis_2019, esscirc_2019} and spin-based semiconductor qubits~\cite{Maurand_2016, Crippa_2018, Corna_2018}, here we discuss the feasibility of realizing a semiconductor charge qubit in the CMOS technology. Our proposed charge semiconductor qubit originates from a fundamental device known as a single-electron device (SED)~\cite{Likharev_1999,Wasshuber01,Ali_2010,Vink_2007}. This device allows to precisely control and manipulate individual electrons. To further facilitate the electron transport and control, the SED can be refined into a single-electron transistor (SET), see Fig.~\ref{fig:SET}. Multiple gate extensions will enable controllable movement of individual electrons as quantum dot as their superposition and entanglement ~\cite{Fujisawa_2004,ICECS18,Giounanlis_2019, esscirc_2019}. A qubit based on the SET is also referred to as a `charge qubit', and was very actively studied between the early 90s and mid-noughties (see, for instance, Ref.~\cite{Gorman_2005}). 

However, concerns about the decoherence time and charge noise~\cite{Petta_2010}, resulted in a period of relative inactivity in this field, compared with other technologies. Very recently, interest in charge qubit devices has resurged, in part because previous problems are now mitigated by new technological advances, especially the fine nanometer-scale feature size of CMOS lithography and short propagational delay afforded by the cut-off frequency ($f_T$) reaching almost a terahertz.

Our research is motivated by recent, significant efforts made to advance quantum qubits and quantum gates implemented in semiconductor and, in particular, CMOS technologies, with a number of very recent studies reporting silicon quantum dots~\cite{Yang_2019,Maurand_2016, Crippa_2018, Corna_2018, Ghosh_2017,Chan_2019} and support electronics~\cite{Charbon_2017, Patra_2018, esscirc_2019,Bluhm_2019, Lehmann_2019, Ekanayake_2010, Pakkiam_2018, Bonen_2018}. These studies demonstrate a more practical view on quantum computing, highlighting the feasibility of large-scale fully integrated quantum processors, where many qubits will be controlled by the means of conventional electronic circuitry. The presented study is particularly relevant to such implementations of quantum processors~\cite{esscirc_2019,Leipold_2019b} which may become dominant architectures in the future.

In support of the scientific and technological feasibility of the proposed charge qubit, the purity of silicon in modern high-volume commercial nanometer-scale CMOS has {\it dramatically} increased since the previous wave of realizations of the charge qubit~\cite{Zong_2015}. This has been driven by increasingly sophisticated CMOS lithographic processes with ultra-high switching speed of devices as well as improved defect tolerances, required to achieve ever increasing densities of functioning transistors (tens of millions per mm$^2$) on microprocessor chips. Hence, the semiconductor qubit under study exploits this highly refined CMOS manufacturing processes, using ultra-pure silicon, precise control of doping, and high control of the interface between silicon and silicon dioxide.

Semiconductor/silicon qubits have been studied both theoretically and experimentally in the literature, taking into account spin, valley and orbital degrees of freedom~\cite{Veldhorst_2015, Sarma_2011}. The manipulation of various quantum states of such systems is based on the control of external electrical or magnetic fields to achieve desired qubit operations~\cite{Li_2018}. Most of these works rely on the manipulation of either spin or both the spin and charge of a particle (hybrid qubits), and usually are restricted to the analysis of a {\it single} double-quantum-dot (DQD) or three quantum-dots~\cite{Sarma_2011,Baart_2016,Ghosh_2017b, Hensgens_2017}.

As aforementioned, a large number of various charge qubit implementations have also been reported in the literature. Briefly outlining some implementations, we note that charge qubits implemented as DQDs based on a Josephson circuit~\cite{Pashkin_2003}, semiconductor charge qubits fabricated in AlaAS/GaAs~\cite{Fujisawa_2011}, a possibly large-scale implementation of charge-based semiconductor quantum computing~\cite{Hollenberg_2004}. Among the most recent studies, Ref.~\cite{Yang_2019} introduces high-fidelity single-qubit gates. Charge qubits in van der Waals heterostructures are discussed in~\cite{lucatto2019charge}. Relevant to charge qubits, an electron localization due to Coulomb repulsion is investigated along with the time evolution of quantum states in the presence of charge noise~\cite{Yang_2019, Chan_2019, Pashkin_2003, Fujisawa_2011, Petersson_2010}.

In this work, we focus our attention on a system based on the electrostatic manipulation of single-electron semiconductor charge qubits implemented in modified structures based on FDSOI 22~nm technology~\cite{esscirc_2019}. We extend the methodology, commonly applied to a double quantum dot system, to the case of multiple-particle qubits, each having an arbitrary number of energy states, interacting electrostatically.

We also investigate entanglement between two or more interacting qubits by the use of the Von Neumann entanglement entropy. Other approaches to measure entanglement between two qubits have been suggested in the literature, for example with the use of a correlator or the concept of concurrence~\cite{Emary_2009, Pashkin_2003_Entanglement}. In this study, we are interested to provide a proof of entanglement between an arbitrary number of qubits or DQDs. However, we should mention that the measurement of entanglement entropy is an open problem and cannot be achieved in a straightforward way, especially for multi-particle systems~\cite{Islam_2015}.

Lastly, we should point out that this study, even though it is focused on charge qubits, can be extended to spin or hybrid qubits, which are currently perceived as the leading trend in CMOS. Specifically, particles with information encoded in spin also need to be moved around with positional precision in a quantum processor. From this point of view, each particle would also undergo, at some point during the computations, some type of shuttling across different QDs to reach another part of the processor and to carry the quantum information. The body of theoretical work on this so-called spin-bus architecture is extensive and has also been reported experimentally~\cite{Baart_2016}. The presented formalism in this study extents the description from a DQD cell to a multiple-QD cell (and even multiple multi-QD cells).

\begin{figure}[t!]
\centering
\includegraphics[width=0.75\columnwidth]{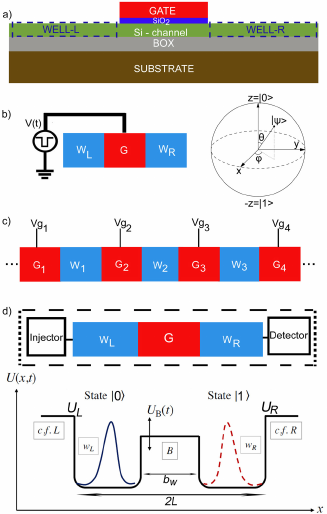}
\caption{(a) A representative example of a CMOS transistor-like silicon-on-insulator device that serves as a coupled quantum-quantum dot system. (b) Charge qubit formed by two coupled quantum dots, each of which acting as a quantum dot (QD). (c) QDs in series forming a quantum register. (d) Block diagram of the system showing an injector and a detector. The injection of an electron is performed through an injector on the left quantum dot of the register whilst the readout is carried out on the right quantum dot by using a single-electron detector. The double-QD (DQD) system forms a charge qubit. The potential function $U(x,t)$ appearing in the Hamiltonian of the system is controlled by the voltages applied at the terminals of the structure, with the barrier $U_B(t)$ varying in time in the most general case. When in a coherent state, an electron injected into such a system can tunnel quantum mechanically through the barrier between the two quantum dots. The electron exists in a superposition of left and right quantum quantum dot states described by its wavefunction. Measurement of the electron position causes wavefunction collapse (it is a destructive/projective measurement); the electron is found to be in either the left or right quantum dot with probabilities related to the wavefunction density (repeated independent measurements yield the left and right position probabilities). Physical device parameters are given in Table 1.}
\label{fig:system}
\vspace{0mm}
\end{figure}

This paper is organized as follows. Section~\ref{sec:statement} presents the statement of the problem, describing a chain of transistor-like QD devices implementing a quantum register. Section~III proceeds with the formal development of a charge qubit whose quantum logic states are defined by the detection of an electron in a specific quantum dot of the quantum register. That section provides the rigorous definition of the qubit and shows how to construct maximally localized functions. In addition, we discuss the robustness of the charge qubit and possible methods to control the angles of its Bloch sphere. Section~IV is dedicated to the derivation from first principles of the tight-binding model for this system. We show how to obtain the Hamiltonian matrix elements and solve the relevant equations for one qubit, and go on to extend the procedure for multiple charge qubits interacting electrostatically. Selected cases of interest, and comparison of the tight-binding formalism with the Schr\"odinger equation, are presented.
Finally, Section~V defines the Von Neumann entanglement entropy $S_{\text{N}}$ in terms of the reduced density matrix in the context of charge qubits. This is used to demonstrate entanglement between states and the importance of the Coulomb electrostatic interaction on the entanglement of charge qubits.

\section{Description of the System Under Study}

\label{sec:statement}

The key building block of the proposed semiconductor charge qubit can be realized in CMOS fully depleted silicon-on-insulator (FDSOI) technology~\cite{Trabesinger_2017} and is shown in Figs.~\ref{fig:system}(a) and (b). It resembles a transistor and comprises two depleted silicon dots separated by a silicon channel, which acts as a tunneling barrier whose potential energy is controlled electrostatically by the gate. Each dot acts as a single quantum dot (QD). When the barrier separating the QDs is very high, quantum mechanical tunneling is exponentially suppressed and the QDs are effectively decoupled. A single electron injected into the system is then trapped in either left or right dot and the quantum state has a very long lifetime. By lowering the barrier, a single electron can tunnel between the left and right dots in the double QD (DQD) device. The potential barrier $U_B(t)$ between the two dots, controlled by the voltage applied at the gate of the device, can vary with time in the most general case, and hence allows control over the electronic tunnelling in the DQD (see Fig.~\ref{fig:system}(d)). To complete the structure, one adds an injector (a device which is able to inject a single electron into one quantum dot) and a detector (a device which is able to detect an electron at the same or the other quantum dot). This geometry allows one to define a charge qubit. We assume that the state of the qubit, as a closed system, can be expressed as a superposition of eigenstates.
\begin{figure}[t]
    \centering
    \includegraphics[width=\columnwidth]{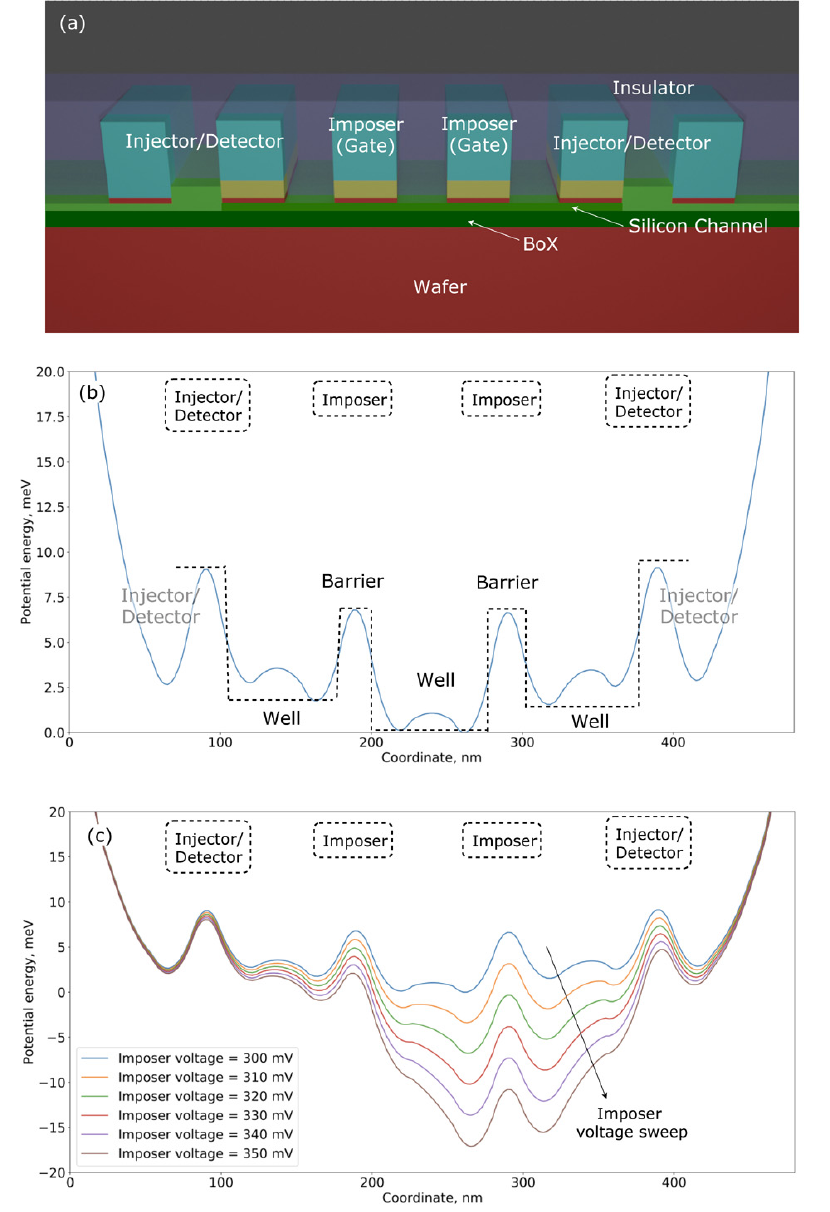}
    \caption{(a) Schematic 3D structure containing gates (imposers), silicon channel, buried oxide (BoX) and wafer. (b)Finite element method (FEM) simulations of the electrostatically shaped potential energy as a function of the coordinate along the structure. By manipulating the electric potential applied at the gates, one can achieve desired potential energy profiles; It can be seen that the potential energy can be approximated by an equivalent piece-wise linear function. The charge qubit can be defined using two potential energy wells separated by a barrier. (c) One can achieve a desired potential energy profile along the coordinate to facilitate or decrease tunneling of an electron between adjacent dots. }
    \label{fig:COMSOL_potential_profiles}
\end{figure}

As this study aims to support the design of a quantum processor in FDSOI 22-nm CMOS technology, finite-element method (FEM) simulations of 2D and 3D structures have been carried out using semiconductor, electromagnetic and Schr\"odinger-Poisson simulators of COMSOL Multiphysics using the dimensions, materials and dopant concentrations of that technology over temperatures 2--70~K to support the model presented in this study. From these simulations, the potential energy of an electron is calculated. The potential energy on the surface of the silicon channel along the symmetry line is then used in the tight-biding model. Both the semiconductor simulation of the modulation of the conduction band by the applied electric field and electromagnetic simulators (the penetration of the electric field in the channel) return consistent results, showing the freeze-out of the channel and the depth of the quantum wells forming along the structure and controlled by the potential applied at the gates (imposers). The schematic 3D structure is shown in Fig.~\ref{fig:COMSOL_potential_profiles}(a) and it contains gates (imposers) made of a stack of SiO$_2$, high-$\varepsilon$ dielectric and heavily doped polysilicon, thin silicon channel, buried oxide (BoX) and thick wafer. An additional insulating coat is deposited on the top of the structure. Some minor effects, such as trapped gate charges, are also taken into account. At the beginning (and also at the end) of the structure, the devices serving as injectors/detectors are connected.

A representative example of the electron's potential energy on the surface of the silicon channel of a quantum register with three dots, as obtained from FEM simulations, is shown in Fig.~\ref{fig:COMSOL_potential_profiles}(b). In such a geometry, there exists such a combination of gate voltages that causes potential energy ``barriers'' to be formed under the imposers and ``wells'' to be formed in between the imposers. The minimum of the potential energy is conventionally placed at 0~meV. In a typical scenario, barriers of 2 to 4~meV are formed when a sub-threshold voltage is applied at the imposers.
The resulting potential energy can be effectively approximated by an equivalent piece-wise linear function. The Schr\"odinger equation with a piece-wise linear potential energy can be solved to find a set of eigenfunctions and eigenenergies. We use three different methods to ensure that the eigenenergies are consistent. Any DQD in a register can be viewed a charge-based qubit. However, the term position-based qubit has also been used~\cite{Harrison_2016, Hollenberg_2004}.

The electron's potential energy can be controlled by applying appropriate potential at the imposer terminals. The imposer voltages allow DC bias voltage and pulses of individually controlled magnitude and duration to be applied to control the barrier height separating the pairs of neighboring DQDs. Figure~\ref{fig:COMSOL_potential_profiles}(c) shows how the potential energy changes when we sweep the potential at one of the imposer over the range from 300 to 350~meV. One can note that the relative height of the barrier separating dot~2 and dot~3 decreases, facilitating tunneling between the two quantum dots.

\section{Formal Definition and Electrostatic Bloch Sphere Control of a Charge Qubit}\label{sec:qubit}

The parameters of our investigated DQD system are given in Table~\ref{tab:params} and correspond to the device shown in Fig.~\ref{fig:system}(d). Here, $e$ and $m_e$ denote the electron charge and mass, respectively, $2 L$ is the length of the DQD device, $b_w$ is the length of the barrier separating the two quantum dots, and $U_B$ is the barrier potential. The left and right dots themselves are at potential $U_L$ and $U_R$. These chosen parameters correspond to 22-nm FDSOI CMOS. For the sake of completeness, all variables are concentrated in one table, although $x_d$ and $z_d$ will be later defined in Fig.\,\ref{fig:3dotsystem} for separated DQD structures.
When coupled to other similar QDs in a chain, as shown in~Fig.~\ref{fig:system}(c), the dynamics of an injected electron can be manipulated on a larger scale. Such an array of QDs is similar to a charge-coupled device (CCD) and allows formation of a quantum register.

\begin{table}[h]
\caption{System parameters for a two-quantum dot qubit(s).}
\centering
\begin{tabular}{|l|l|}
\hline
 $e$ & $1.6\cdot 10^{-19}$ C \\
 \hline
 $m_e^*$ & 0.2$\cdot$ $m_0$ \quad \text {and} \quad 1.08$\cdot$ $m_0$ \\
 \hline
 $L$ & $80$ nm \\
 \hline
 $b_w$ & $0.25\,L$\\
 \hline
 $E_0$ & $\hbar^2/2 m_e^{*} L^2$ = 0.006meV \\
 \hline
 $U_L=U_R$ & $300\, E_0$= 1.8 meV\\
 \hline
 $U_B$ & $120\, E_0$ = 0.72 meV \\
 \hline
 $x_d, \,z_d$ & $2\, L$ \\
 \hline
\end{tabular}
\label{tab:params}
\end{table}

A qubit is generally defined as an isolated quantum system that has two distinctive quantum states (denoted $\ket 0$ and $\ket 1$) controlled via various technical means (for example, by applying electric or/and magnetic fields). In the case of a qubit implemented through two coupled quantum quantum dots (i.e. a DQD), electrostatic or electromagnetic fields facilitate occupancy (Rabi) oscillations between the two states. The spectral theorem guarantees that two different eigenvalues of the system's Hamiltonian, expressed as a Hermitian matrix, have orthogonal normalized eigenstates, which is essential for the operation of a qubit~\cite{Tannor_2007}. The fidelity is preserved provided the quantum system remains effectively isolated from any decohering environment over the timescales of the experiment. In this section, we aim to show the feasibility of charge qubits and introduce their formal definition.
To analyze the eigenstates of a two-dot or multi-dot system, we begin with the time-independent Schr\"odinger formalism. Later we will show an extension to a time-dependant Hamiltonian and multi-particle case. The Schr\"odinger equation is written as follows:
\begin{equation}
\hat{\text H} \ket{\psi_j(x)} = E_j \ket{\psi_j(x)} \;
\label{eq:1}
\end{equation}
where $\hat{\text H}$ is the Hamiltonian operator for the system, $\ket{\psi_j(x)}$ is an eigenstate labelled by index $j$, and $E_j$ is its corresponding energy. For a time-independent Hamiltonian, the wavefunction dynamics can be obtained in the Schr\"odinger picture simply from $\ket {\Psi_j (x,t)} = e^{-i E_j t/\hbar}\ket{\psi_j(x)}$, where $i$ is the imaginary unit.

Consider the simplest case with only two energy levels, $E_0$ with corresponding wavefunction $\ket{\psi_0}$, and $E_1$ with wavefunction $\ket{\psi_1}$. At any given time $t$, the state of a qubit can be represented in terms of the superposition,
\begin{equation}\label{eq:qstate}
\ket \psi = c_0 \ket {\psi_0} + c_1 \ket {\psi_1}
\end{equation}
where $c_0$, $c_1$ are the probability amplitudes of each eigen-state in the $\ket \psi$ basis, with $|\alpha|^2+|\beta|^2 = 1$ to preserve the normalization of the wavefunction.

However, the eigenstates $\ket{\psi_{0}}$ or $\ket{\psi_{1}}$ do not typically correspond to states of a single electron in a DQD that are physically detectable by means of a standard electrostatic (charge) detector, as used in the proposed device. Instead, we change the basis and write $\ket \psi$ in terms of the \textit{detectable qubit states} $\ket{0}$ and $\ket{1}$ \cite{Giounanlis_2019},
\begin{equation}\label{eq:LRbasis}
\ket{\psi}=c_0\ket{0} + c_1\ket{1} \equiv \cos\tfrac{\theta}{2}\ket{0} + e^{i \varphi} \sin\tfrac{\theta}{2}\ket{1}
\end{equation}
where $|c_0|^2 + |c_1|^2 = 1$, and the angles $\varphi \in [0,\,2\pi)$ and $\theta \in [0,\pi]$ define the so-called Bloch sphere representation of a qubit. Since the QDs are physical quantum quantum dots with spatial extent along the lateral $x$-axis as shown in Fig.~\ref{fig:system}(d), the states $\ket{0}$ and $\ket{1}$ are associated with time-independent wavefunctions $\ket{\phi_L (x)}$ and $\ket{\phi_R (x)}$, defined such that $\ket{\phi_L (x)}$ maximizes the electronic occupancy of the left quantum dot and $\ket{\phi_R (x)}$ maximizes the electronic occupancy of the right quantum dot. A prescription to determine these functions is given below.
First, note that the coefficients $a$ and $b$ are generally complex-valued, following from orthonormality as,
\begin{equation}\label{eq:LRcoef}
c_0 = \braket{0}{\psi} \triangleq \int\limits_{\mathcal D}^{}\phi_L^*\psi\, \text d x \,\,\, \text{and} \,\,\, c_1 = \braket{1}{\psi} \triangleq \int\limits_{\mathcal D}^{}\phi_R^*\psi\, \text d x
\end{equation}
where $x \in \mathcal D$ denotes the entire domain of existence of an electron injected into the two-quantum dot system with finite walls. Hence, Eqs.~\eqref{eq:LRbasis}--\eqref{eq:LRcoef} define formally a charge qubit. We emphasize that the new orthonormal states $\ket{\phi_L(x)}$ and $\ket{\phi_R(x)}$ are a linear combination of the original eigenstates $\ket{\psi_0(x)}$ and $\ket{\psi_1(x)}$. This is not surprising since the physical system considered has only two energy levels, and so only two orthogonal wavefunctions are available to construct the new basis. This formalism can be straightforwardly generalized to multi-quantum dot and multi-level system.

In general, the basis transformation is linear since the Schr\"odinger equation is a linear differential equation, and takes the form,
\begin{equation}\label{eq:position_transform}
\ket{\phi_{\zeta=L,R}}=\sum_{j=0,1} U_{\zeta j}\ket{\psi_j}
\end{equation}
where $U_{\zeta j}=[\hat {\text U}]_{\zeta j}$ are elements of a unitary matrix $\hat {\text U}$. Since $\hat{{\text U}^{\dag}}\hat{\text U}=\hat {\text I}$ for a unitary matrix and the original eigenstates satisfy $\langle\psi_i|\psi_j\rangle=\delta_{ij}$, the states of the new basis are guaranteed to be orthonormal and the antisymmetry of the fermionic wavefunction is preserved (this is referred to as a `canonical transformation'). Here, the dagger symbol denotes Hermitian conjugation, and $\delta_{ij}$ is the Kronecker delta symbol.

Of course, there exist an infinite number of unitary matrices $\hat{\text U}$ that satisfy Eq.~\eqref{eq:position_transform}. To complete the definition of the charge qubit, we must find the specific representation that satisfies an additional constraint -- namely, that $\ket{\phi_L(x)}$ maximizes the occupancy of the left quantum dot, and $\ket{\phi_R(x)}$ maximizes the occupancy of the right quantum dot.

It should be emphasized here that to faithfully model the physical DQD device, the quantum quantum dots do not have infinite potential walls. Although the probability of locating the electron in either quantum dot is relatively high, there is a finite probability that the electron can exist in classically forbidden regions, such as in the barrier region between the quantum dots, or outside of the device entirely, leading to a loss of quantum information from the system. This motivates us to define the ``robustness'' of the charge qubit and to estimate its non-ideality. In particular, we wish to construct a basis that minimizes any such non-ideality.

We start by recalling that the total probability of locating an injected electron across the entire domain of existence is exactly unity (this is the physical property responsible for wavefunction normalization). Therefore,
\begin{equation}
\begin{split}
\int\limits_{\mathcal D}^{}|\psi|^2 \,\text{d}x = & \int\limits_{\text{c.f.L \& c.f.R}}^{}|\psi|^2 \,\text{d}x +
\int\limits_{\text{B}}^{}|\psi|^2 \,\text{d}x + \\ &\int\limits_{w_L}^{}|\psi|^2 \,\text{d}x + \int\limits_{w_R}^{}|\psi|^2 \,\text{d}x = 1
\end{split}
\end{equation}
where `c.f.L' and `c.f.R' stand for the classically forbidden regions outside the left and right quantum dots, `B' stands for the controllable barrier region separating the two quantum dots, and $w_L$ and $w_R$ are respectively the left and right quantum dots themselves (see Fig.~\ref{fig:system}(d)).

We define the probability of locating an electron in quantum dot $\zeta=L$ or $R$ as $p_{w_{\zeta}}$. From Eq.~\eqref{eq:LRbasis} it then follows that,
\begin{equation}\label{eq:fdef}
p_{w_{\zeta}}= \int\limits_{w_{\zeta}}^{}\left( |c_0|^2 |\phi_L|^2 + |c_1|^2 |\phi_R|^2 + c_0\,c_1^*\phi_L\,\phi_R^* + \text {c.c.} \right)\,\text{d}x \;
\end{equation}
where `c.c' denotes complex conjugate. We recognize that $\phi_L(x)$ and $\phi_R(x)$ are not vanishing at the barriers but have (exponentially) decaying tails that constitute the dominant source of non-ideality of the charge qubit. A measure of the non-ideality (or residual error factor) is then $\epsilon=1-p_{w_{L}}-p_{w_{R}}$.

The localized-state basis of \textit{Wannier functions}~\cite{Marzari_1997,Mostofi_2008} is defined such as to maximize the probability to locate an electron in the relevant quantum dot. Using formula~\eqref{eq:position_transform} for $\zeta = L$~or~$R$, we write:
\begin{equation}
\begin{split}
&\{(x,\phi_L(x)):x\in \mathcal D\} \quad \text{and} \quad \{(x,\phi_R(x)):x\in \mathcal D\} \\
&\text{max}\int\limits_{w_\zeta} |\phi_\zeta|^2 dx = \\ &\text{max} \int\limits_{w_\zeta} (U_{\zeta 0} \cdot \psi_0 + U_{\zeta 1} \cdot \psi_1)(U^*_{\zeta 0} \cdot \psi^*_0 + U^*_{\zeta 1} \cdot \psi^*_1)dx\\
\end{split}
\end{equation}
This optimization problem allows one to find the matrix elements $U_{\zeta j}$ performing the basis transform.

\begin{figure}[t!]
\includegraphics[width=\columnwidth]{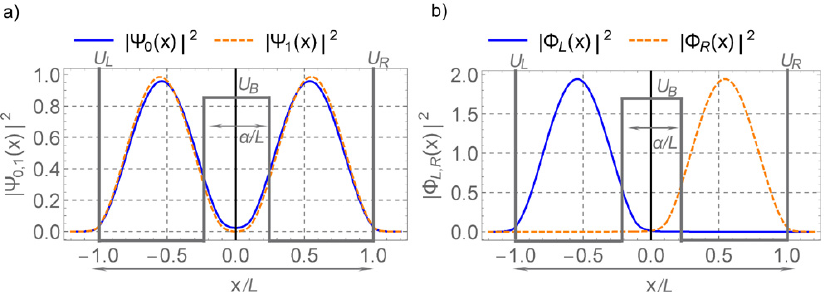}
\centering
\caption{Representation of qubit states in the proposed DQD device. (a) Probability density of and electron in eigenstates of the DQD, $|\ket{\psi_0(x)}|^2$ and $|\ket{\psi_1(x)}|^2$ as a function of position $x/L$. (b) Corresponding maximally localized functions $|\ket{\phi_L (x)}|^2$ and $|\ket{\phi_R(x)}|^2$.}
\label{fig:basis}
\end{figure}

For a DQD qubit comprising two quantum quantum dots with the parameters of Table~\ref{tab:params}, one can straightforwardly determine the maximally localized basis functions, as shown in Fig.~\ref{fig:basis}. Figure~\ref{fig:basis}(a) shows the probability density for an electron in eigenstate $\ket{\psi_0(x)}$ or $\ket{\psi_1(x)}$ as a function of position along the $x$-axis, highlighting how the eigenbasis is typically delocalized over the entire device. By contrast, Fig.~\ref{fig:basis}(b) shows the probability density for an electron in the maximally localized (Wannier) basis $\ket{\phi_L (x)}$ or $\ket{\phi_R (x)}$, demonstrating suitability as a charge qubit basis. Note however that even in the maximally localized basis, there is appreciable tunneling amplitude inside the classically forbidden barrier region.

\begin{figure}[b!]
\includegraphics[width=\columnwidth]{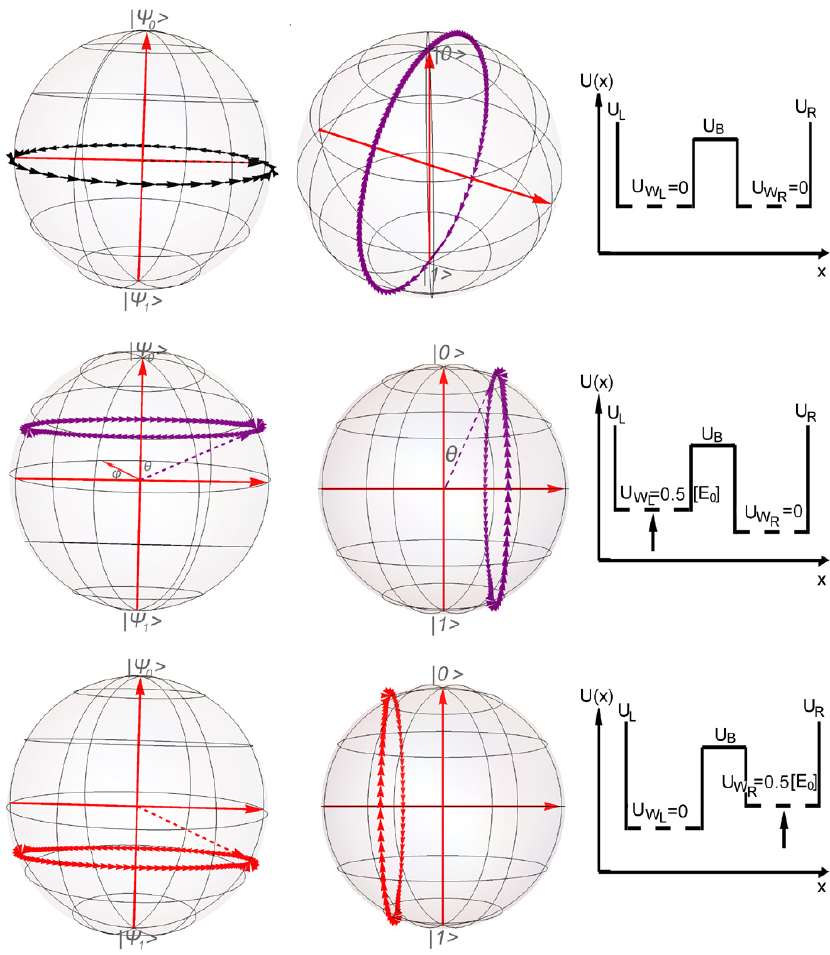}
\centering
\caption{Rotation of angle $\theta$ both in the eigenfunction basis $\{\ket{\psi_0},\ket{\psi_1}\}$ and position basis $\{\ket{0},\ket{1}\}$ (i.e. the Bloch sphere). Angle $\theta$ can be adjusted by dynamically manipulating the potential function $U_B(x)$ as quantum dot as by setting the bottoms of the potential quantum dots $U_{w_L}$ and $U_{w_R}$.}
\label{fig:bloch_sphere}
\end{figure}

The full control of the Bloch sphere requires one to be able to change both angles, $\theta$ and $\varphi$. However, it is easy to see that an equilibrium system in the eigenfunction representation is characterised by a fixed angle $\theta$ with the angle $\varphi$ precessing at the frequency of occupancy oscillations $\delta \omega =
(E_1 - E_0)/\hbar$ where the energy levels $E_0$ and $E_1$ are associated with the states $\ket{\psi_0}$ and $\ket{\psi_1}$. Angle $\theta$ can be adjusted by dynamically modulating the potential function $U_B(x)$ as quantum dot adjusting the bottoms of the potential quantum dots $U_{w_L}$ and $U_{w_R}$. However, once the system is in equilibrium (i.e., $U_B(x)$ is high enough), $\theta$ does not change. The eigenfunction representation of the two-quantum dot system is shown in Fig.~\ref{fig:bloch_sphere} (see the left column) where we show the effect of the potential function variation on the angle $\theta$. The state vector describing such a system precession with the frequency $\delta \omega$ along the paths shown in those figures is defined by the height of the barriers separating the quantum dots.



It is interesting to note that in the position-based representation~\eqref{eq:position_transform}, the Bloch sphere and the original trajectories are transformed as shown in the middle column of Fig.~\ref{fig:bloch_sphere}, with both angles, $\theta$ and $\varphi$, being functions of time even in the equilibrium case. It is straightforward to obtain the expressions for the angles in explicit form:
\begin{equation}
\begin{split}
&\cos^2 \frac \theta 2 = \frac 1 2 + |c_0| |c_1| \cos (\delta \omega t) \\
&\sin^2 \frac \theta 2 = \frac 1 2 - |c_0| |c_1| \cos (\delta \omega t)\\
&\varphi = \text{arctan}\left [ \frac {2 |c_0| |c_1| \sin (\delta \omega t)}{|c_0|^2- |c_1|^2}\right]
\end{split}
\end{equation}
%
Here, according to Eq.~\eqref{eq:qstate}$, |c_0|$ and $|c_1|$ represent the probability amplitudes of the eigenstates $\psi_0$ and $\psi_1$, and the frequencies $\omega_0 = E_0 /\hbar$ and $\omega_1 = E_1 /\hbar$ are associated with those states.

\section{Deriving Tight-Binding Model of Interacting Qubits from Schr\"odinger Formalism}\label{sec:tb}

The Schr\"odinger formalism, both time-dependent and time-independent, allows one to capture the dependence of the wavefunction on spatial coordinates from first principles. However, it becomes increasingly inconvenient to use when handling multiple interacting particles or more complex structures. In this section, we show how to derive a tight-binding model directly from the Schr\"odinger equation, allowing us to model single and multiple electrons in such structures easily and effectively. The tight-binding model is often used in systems where localized Wannier orbitals constitute a good basis for quantum tunneling of electrons in a periodic potential~\cite{Vogl_1983, Xu_2002,Schulz_2007, Goldmann_2014}. In bulk materials treated within the tight-binding formalism, the long-range Coulomb interaction is often assumed to be screened, but in quantum confined nanostructures as studied here, the electron-electron interactions are crucial and must be considered explicitly. The aim here is to show how the parameters appearing in the model will be related directly to the geometry of the system and to the maximally localized Wannier functions introduced in the previous section.

\subsection{Multi-Particle Formalism in Application to the Studied System}

For convenience, we will use the first quantization formalism. The Hamiltonian of $N$ interacting particles in one dimension contains the kinetic energy operator $-\sum\tfrac{\hbar^2}{2m^*}\nabla^2_k$, the potential energy operator $\sum U(x_k)$ and the interaction energy $\sum U^C_{kj}$ due to the Coulomb force between electrons. Hence, the Hamiltonian operator is written as follows:
\begin{equation}\label{eq:h-many-particles}
\begin{split}
\hat{\text H}=\sum_{k=1}^{N} \left(- \frac{\hbar^2}{2m^{*}_k}\nabla_k^2 + U_k(x_k,t)\right) +
\sum_{k >j = 1}^N \frac{g\cdot e^2}{ 4\pi\varepsilon_{\text{eff}}|x_k - x_j|}
\end{split}
\end{equation}
where $\varepsilon_{\text{eff}}$ is the effective dielectric constant and $g$ is a coefficient accounting for screening effects~\cite{Kruchinin_2010}, $e$ is the magnitude of the electronic charge and $m^{*}_k$ is the (effective) mass of the $k^{\text{th}}$ particle.
 It must be noted that if equation~\eqref{eq:h-many-particles} is used in a one-dimensional case, the correction for dimensionality for the electric potential should be taken into account.
The time-dependent Schr\"odinger equation is then written in terms of the $N$-particle eigenstate wavefunctions $\ket {\Psi_j}\equiv \ket {\Psi_j(x_1,\ldots, x_N,t)}$ as follows:
\begin{equation}\label{eq:tdse}
i \hbar \frac{\partial \ket {\Psi_j}}{\partial t} = \hat {\text H} \ket{\Psi_j}
\end{equation}

As usual, $N$-particle wavefunctions $\ket {\Psi(x_1,\ldots, x_N,t)}$ can be represented in terms of a linear combination of $N$-particle basis states $\ket{\psi(x_1,\ldots, x_N,t)}$, which are themselves constructed as the tensor product of the $M$ single-particle states of each particle $i$, $\ket{\psi(x_i,t)}=\sum_{n_i=1}^M c_{n_i}^{(i)} (t) \ket{\phi_{n_i}^{(i)}(x_i)} $, according to:
\begin{equation}\label{eq:wavefunc}
\begin{split}
&\ket{\psi(x_1, \ldots, x_N,t)} = \ket{\psi(x_1,t)}\otimes \ldots \otimes \ket {\psi (x_N,t)} = \\
&\sum_{n_1=1}^M c_{n_1}^{(1)} (t) \ket{\phi_{n_1}^{(1)}(x_1)} \otimes...\otimes \sum_{n_N=1}^M c_{n_N}^{(N)}(t) \ket {\phi_{n_N}^{(N)}(x_N)}
\end{split}
\end{equation}
In the combined Hilbert space $\cal{H}^{(\otimes N)}$, we introduce the basis $\ket{\phi^{(\otimes N)}}$:
\begin{equation}\label{eq:basis}
\begin{split}
&\left\{\{\phi_k^{(\otimes N)}\}:\;\ket{\phi_{n_1}^{(1)},...,\phi_{n_N}^{(N)}} =\right. \\ &= \left.\ket{\phi_{n_1}^{(1)}}\otimes\ldots\otimes \ket{\phi_{n_N}^{(N)}}, n_1,..,n_N = 1,...,M \right\}
\end{split}
\end{equation}
where $k=1,..., {M^N}$ with $M^N$ providing the total number of such basis functions. The Schmidt decomposition theorem states that all states in the combined Hilbert space can be expressed as a linear combination of these tensor product states, so we write:
\begin{equation}\label{eq:sum-wavefunc}
\begin{split}
\ket {\Psi} = \sum_{k=1}^{M^N}c_k \ket{\phi_k^{(\otimes N)}}
\end{split}
\end{equation}

In the basis of $\ket{\phi_k^{(\otimes N)}}$, the Hamiltonian matrix is not diagonal, and so in general we have finite matrix elements of the type
\begin{equation}\label{eq:h-elements}
H_{mn}=\bra{\phi^{(\otimes N)}_m} \hat {\text H} \ket {\phi_n^{(\otimes N)}}
\end{equation}
%
Eq.~\eqref{eq:tdse} then implies the following equation for the time-evolution of the coefficients,
\begin{equation}
i\hbar \frac{\text{d} c_m}{\text{d} t}= \sum_n H_{m n} c_n
\label{eq:System_of_equations}
\end{equation}
Equation~\eqref{eq:System_of_equations} is a linear system of ordinary differential equations, which can be solved analytically for a time-independent Hamiltonian or numerically for a time-dependent commuting Hamiltonian:
\begin{equation}
\mathbf{c}(t)= \mathbf{c_0} e^{-
\frac{i}{\hbar}\int_{0}^t \hat{\text {H}}(\tau)d\tau}
\label{eq:exp_solutions}
\end{equation}
where $\mathbf{c}(t)$ is a vector containing the probability amplitudes~$c_i$ and $\mathbf{c_0}$ is a vector of initial conditions subjected to the usual normalization constraint.
If one deals with a time-dependent non-commuting Hamiltonian, the Dyson series can be used to calculate it numerically.

In addition to the formalism stated above, the postulates of our model are as follows:
\begin{itemize}
\item[$\square$] An electron, injected into a double-quantum dot cell (this could be straightforwardly extended to a multi-quantum dot arrangement) is confined to that cell, even if it interacts with other electrons.
\item [$\square$] We will consider one or a set of interacting double-quantum dot cells, each containing an electron that can occupy the two lowest energy levels. Hence, in formula~\eqref{eq:wavefunc} we take into account only two basis functions for each electron (for example, maximally localised $\phi_{L}$ and $\phi_{R}$ for each particle).
\item[$\square$] The electron's time-dependent wavefunction becomes $\Psi(x,t) = c_0(t) \phi_L(x) + c_1(t) \phi_R (x)$. When a system of $N$ interacting electrons is considered, their individual wavefunctions are combined using formula~\eqref{eq:wavefunc}.
\end{itemize}

The applications of these postulates can be easily understood by the example of one electron in a double-quantum dot cell, see Fig.~\ref{fig:SEvsTB}. If the electron is actualized in the left quantum dot $w_L$, its state is associated with the maximally localized function $\phi_L$. Hence, the wavefunction of the electron when `firmly' detected in $w_L$ at a given instance of time is $\ket 0 = 1\cdot \phi_L (x) + 0 \cdot \phi_R (x)$. The actualization of the electron in the right quantum dot $w_R$ is associated with the maximally localized function $\phi_R$. Hence, the wavefunction of the electron when `firmly' detected in $w_R$ at a given instance of time is $\ket 1 = 0\cdot \phi_L (x) + 1 \cdot \phi_R (x)$.

Equating the maximally localized functions to the actualization of the electron can be seen as an approximation, but it is held to a high degree of accuracy. Indeed, since the functions $\phi_L$ and $\phi_R$ are the solution to the maximization problem~\eqref{eq:fdef}, the instantaneous probability of locating the electron, for instance, in $w_L$ is
\begin{equation}
\begin{split}
p_{w_L} = \int\limits_{w_L}[c_0^* \phi_L^* + c_1^* \phi_R^*][c_0 \phi_L + c_1 \phi_R] \text{d}x \approx \\ |c_0|^2\,\int\limits_{w_L} \phi_L^* \phi_L \, \text{d}x \hspace{0mm} \approx |c_0|^2\, \int\limits_{\mathcal D} \phi_L^* \phi_L \, \text{d}x = |c_0|^2
\end{split}
\end{equation}
which is the probability amplitude of $\phi_L$.

In the next section, we consider the application of this general formalism to three cases of interest. The underlying feature of the studied system is that electrons, while interacting through Coulomb force, stay confined to their respective double-quantum dot cells. We will see that combining the equations obtained from first principles with the postulates we formulated is essentially the \textit{tight-binding} model.

\subsection{One Electron in a Double-Quantum-Dot: Qubit}

\begin{figure}[b!]
\includegraphics[width=\columnwidth]{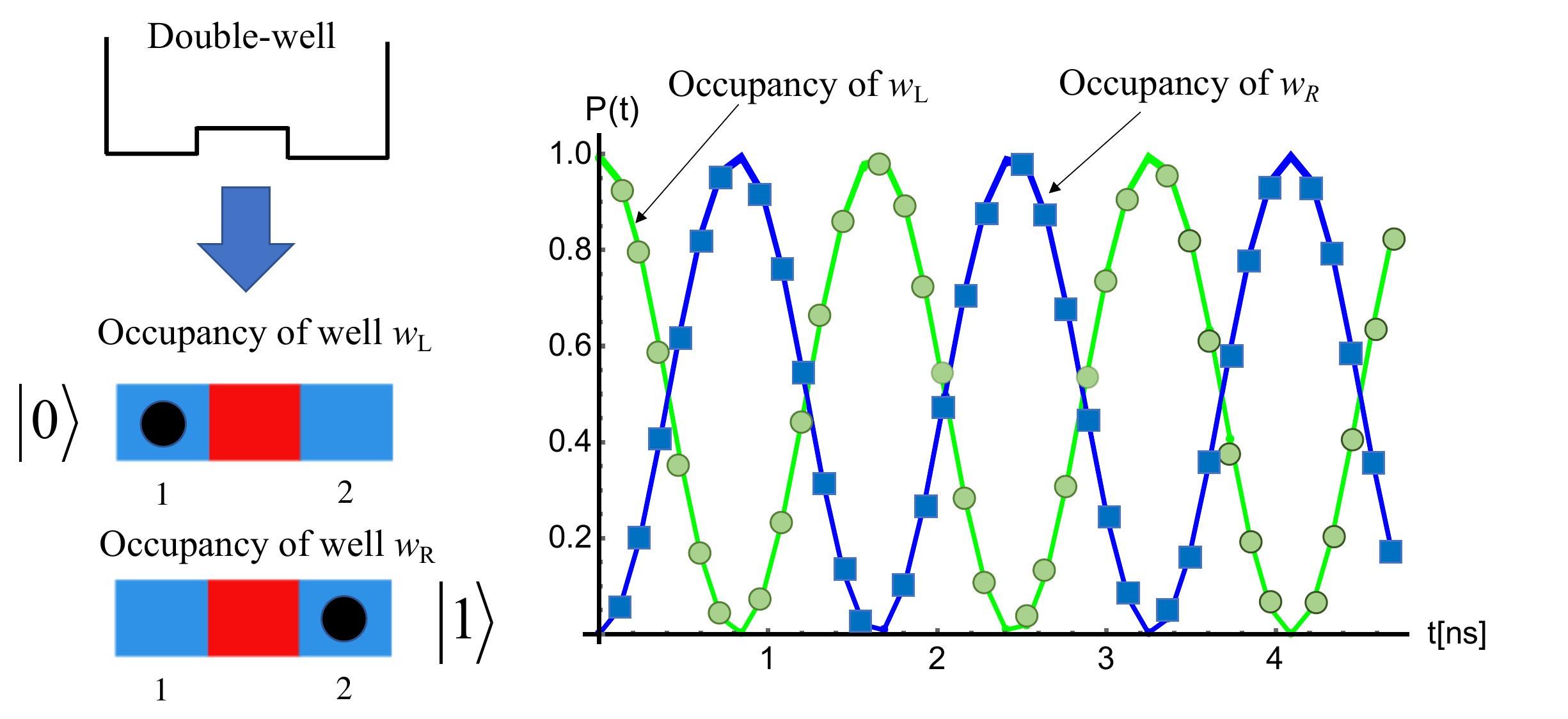}
\centering
\caption{Charge qubit representation: double-quantum dot potential is seen as a cell where the actualisation of the electron in quantum dot $w_L$ corresponds to state $\ket 0$ and the actualisation of the electron in quantum dot $w_R$ corresponds to $\ket 1$. Such a qubit will display the occupancy oscillations between the quantum dots reflected in oscillating probabilities $P_{w_L}(t)$ and $P_{w_R} (t)$ as functions of time. The continuous lines show the probabilities calculated using the equation~\eqref{eq:System_of_equations} with the matrix elements~\eqref{eq:Hcoef} while the squares and circles show the probabilities calculated directly from the Schr\"odinger equation.}
\label{fig:SEvsTB}
\end{figure}

We return to the basic definition of the qubit introduced in Section~\ref{sec:qubit} and analyze it using the framework proposed in Section~\ref{sec:tb}-A with all expressions simplified for the one-particle case. The charge qubit is shown in Fig.~\ref{fig:SEvsTB} where the double-quantum dot structure, the building block of the charge qubit, is represented as a symbolic cell with an electron actualized either in the left or in the right quantum dot. We have already calculated the maximally localized Wannier functions for this system, $\phi_L (x)$ and $\phi_R (x)$, see Fig.~\ref{fig:basis}. The Hamiltonian in the matrix form becomes (not including the Coulomb interaction for the moment since we deal with one electron):
\begin{equation}\label{eq:Hcoef}
\begin{split}
H_{11} &= \bra{\phi_R}\hat {\text H} \ket {\phi_R} = -\frac{\hbar^2}{2m^*}\braket{\phi_R}{\phi_R''} + \bra{\phi_R} U(x) \ket {\phi_R}\\
H_{22} &= \bra{\phi_L}\hat {\text H} \ket {\phi_L} = -\frac{\hbar^2}{2m^*}\braket{\phi_L}{\phi_L''} + \bra{\phi_L} U(x) \ket {\phi_L}\\
H_{12} &= \bra{\phi_R}\hat {\text H} \ket {\phi_L} = -\frac{\hbar^2}{2m^*}\braket{\phi_R}{\phi_L''} + \bra{\phi_R} U(x) \ket {\phi_L} \\
H_{21} &= \bra{\phi_L}\hat {\text H} \ket {\phi_R} = -\frac{\hbar^2}{2m^*}\braket{\phi_L}{\phi_R''} + \bra{\phi_L} U(x) \ket {\phi_R}
\end{split}
\end{equation}
where the double apostrophe symbol denotes the second derivative with respect to coordinate. Hence, knowing the system's geometry and the potential function $U(x)$, it is possible to calculate the functions $\phi_{L,R}$ and then the matrix elements. The calculation of the localized functions $\phi_{L,R}$ is required only once in the case of time-independent system. In the case of weakly perturbed systems, one can assume that the maximally localised functions are not `disturbed' too much, and the time-dependent dynamics are expressed in the probability amplitudes. We note that in the case of cells with symmetrical functions $U(x)$, as the one shown in Fig.~\ref{fig:system}(d), $H_{12} = H_{21}$. 
As an illustration, the matrix entries calculated for our particular geometry are given in Table~\ref{tab:matrix2}. For convenience, we normalize the energy of electrons by the quantity $E_0 = \hbar^2/(2m_e^* L^2)$, so the entries of the Hamiltonian matrix are expressed in units of $E_0$.

\begin{table}[h]
\caption{System parameters for a two-quantum dot qubit.}
\centering
\begin{tabular}{|l|l|}
\hline
 $H_{11} = 9.4\cdot E_0$ & $H_{12} = 0.23\cdot E_0$ \\
 \hline
 $H_{21} = 0.23\cdot E_0$ & $H_{22} = 9.4\cdot E_0$ \\
 \hline
 \end{tabular}
\label{tab:matrix2}
\end{table}

Hence, for a two-level system of Fig.~\ref{fig:SEvsTB} representing the electrostatic qubit, we have the following model. The state when the electron is actualised in $w_L$ is denoted as $\ket{0} \equiv \ket {\phi_L}$ while the state when it is actualized in the right quantum dot --- as $\ket{1} \equiv \ket {\phi_R}$. The time evolution of the states is described by equation~\eqref{eq:System_of_equations} written in terms of the probability amplitudes $\mathbf{c} = (c_1, c_0)^{\text T}$:
\begin{equation}\label{eq:onevector}
\ket{\Psi} = c_1 \ket{1} + c_0 \ket{0}
\end{equation}
The Hamiltonian matrix in that equation becomes:
\begin{equation}\label{eq:Hmatrix}
\textbf{H} = \begin{bmatrix}
 E_{p_1} & t_{h,01}\\
 t_{h,10} & E_{p_2}
 \end{bmatrix}
\end{equation}
It is conventional to use the following notation $H_{11} = E_{p1}$, $H_{22} = E_{p2}$, $H_{12} = t_{h,01}$ and $H_{21} = t_{h,10}$. The off-diagonal terms $t_{h,01}$ and $t_{h,10}$ are known as tunnelling or hopping terms. In this form, the charge qubit is no different than any other quantum two-state system, and hence it will display all the expected features. We note that the matrix~\eqref{eq:Hmatrix} is the fundamental building block for the Hamiltonian matrix of many-particle systems as will be shown later.

Figure~\ref{fig:SEvsTB} plots the electron occupancy oscillation curves in quantum dot $w_L$ (green curve) and quantum dot $w_R$ (blue curve) in the double-quantum dot structure. We used the parameters from Table~\ref{tab:params} and expressions~\eqref{eq:Hcoef} to calculate the entries of matrix~\eqref{eq:Hmatrix}. Then, the set of equations~\eqref{eq:System_of_equations} was solved to find the coefficients $c_0(t)$ and $c_1(t)$ and the probabilities $|c_0|^2$ and $|c_1|^2$ associated with the occupancy of the quantum dots. For comparison, the direct solution of the Schr\"odinger equation is marked by circles and squares, and it provides exactly the same probabilities. As expected, the occupancy of $w_L$ and $w_R$ are in anti-phase. In the case of one electron, it can be localized only in one quantum dot. Hence, when the occupancy of $w_L$ reaches unity, the occupancy of $w_R$ must be zero. In other words, the normalization condition is preserved.

\subsection{Two Interacting Qubits}\label{sec:two-qubits} 

\begin{figure}[t!]
\includegraphics[width=\columnwidth]{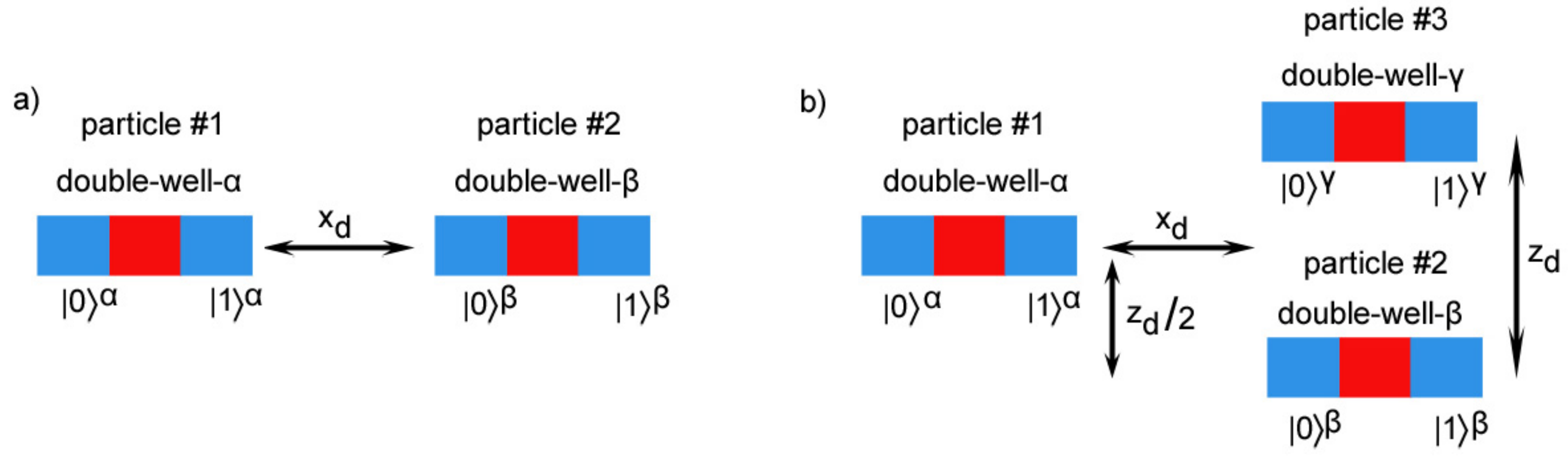}
\centering
\caption{Schematic structures of two and three interacting qubits (i.e., double-quantum dot cells with an electron injected in each of the cells).} 
\label{fig:3dotsystem}
\end{figure}

Now we examine the extension of the formalism to many-particle systems where different DQDs are only electrostatically coupled, with no wavefunction overlap, and where dissipation is not considered. As an illustration, we will study in detail two and three interacting electrons, as illustrated in Fig.~\ref{fig:3dotsystem}. We will use the superscript $\alpha$ to denote the first double-quantum dot cell where particle~1 is injected and the superscript~$\beta$ to denote the second double-quantum dot cell where particle~2 is injected.
For two electrons interacting through Coulomb force, each confined to their respective double-quantum dots, the wavefunction is written as:
\begin{equation}\label{eq:2wavefunc}
\begin{split}
\ket{\Psi}&= \sum_{n_\alpha=1,0}\:\:\sum_{n_\beta=1,0} c_{n_\alpha n_\beta}\ket{n_\alpha^{(\alpha)}n_\beta^{(\beta)}} = \\ &=c_{11} \ket{1^\alpha 1^\beta} + c_{10} \ket{1^\alpha 0^\beta} + c_{01} \ket{0^\alpha 1^\beta} + c_{00} \ket{0^\alpha 0^{\beta}}
\end{split}
\end{equation}
where the states $\ket{0^\alpha}$, $\ket{1^\alpha}$, $\ket{0^\beta}$ and $\ket{1^\beta}$ are associated with the maximally localised functions $\phi_L^{\alpha}$, $\phi_R^\alpha$, $\phi_L^\beta$ and $\phi_R^\beta$, respectively.

\begin{figure}[t!]
\includegraphics[width=\columnwidth]{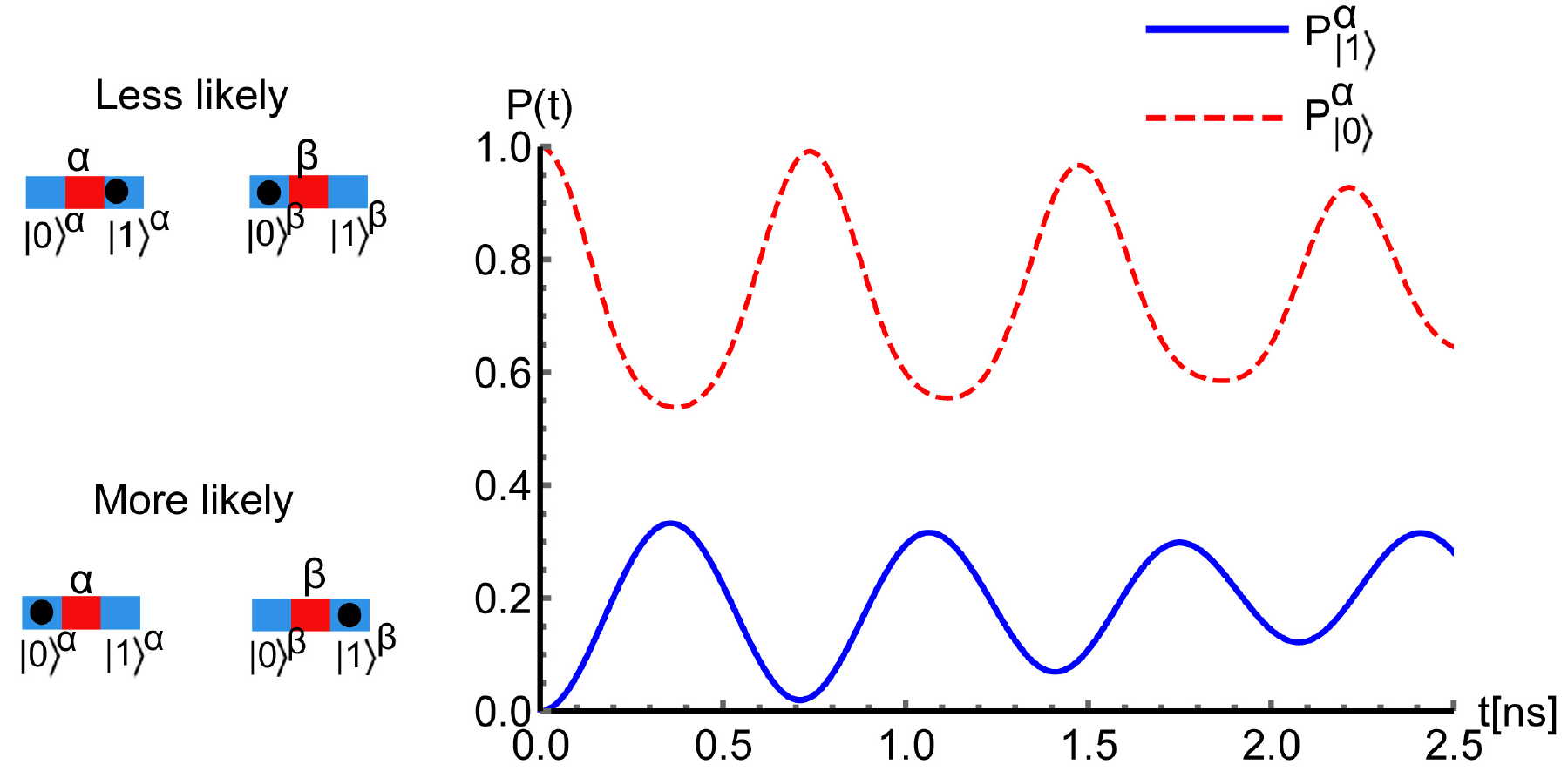}
\centering
\caption{Two interacting qubits, each containing a single electron: occupancy oscillations evolving in time of the electron in double-quantum dot $\alpha$ expressed as probabilities $P^{\alpha}_{\ket{1}}$ and $P^{\alpha}_{\ket{0}}$ to locate the particle in the left or in the right quantum dot in the presence of interaction with the electron in double-quantum dot~$\beta$. The scheme on the left shows the least and most probable states of the particles.}
\label{fig:2dotsystem_evolution}
\end{figure}

The wavefunction contains possible detectable states of the two-particle system. For instance, the probability of finding particle $\alpha$ in the right quantum dot \textit{and} particle $\beta$ also in the right quantum dot of their respective cells is $|c_{11}|^2$, etc.
The Hamiltonian matrix accommodates the electrostatic interaction by including the terms due to the electrostatic interaction and tunnelling terms for both particles by combining formulae~\eqref{eq:h-many-particles} and~\eqref{eq:h-elements}.
%
%
\begin{equation}
\mathrm{ {\textbf H}} = {\textbf H}^{(\beta)}\otimes {\textbf I} + {\textbf I} \otimes {\textbf H}^{(\alpha)} = \begin{bmatrix}
  E_{p_{11}} & t_{h,10}^\beta & t_{h,10}^\alpha & 0\\
  t_{h,01}^\beta & E_{p_{22}} & 0 & t_{h,10}^\alpha\\
  t_{h,01}^\alpha & 0 & E_{p_{33}} & t_{h,10}^\beta\\
  0 & t_{h,01}^\alpha & t_{h,01}^\beta & E_{p_{44}}
  \end{bmatrix}
\end{equation}
where $ {\textbf H}^{\alpha}$ and $ {\textbf H}^{\beta}$ denote the matrices including the Coulomb interaction for the first and second particles and $ {\textbf I}$ is the $2\times 2$ identity matrix.

System of equations~\eqref{eq:System_of_equations}, solved for the two-particle case, allows one to look at different configurations of electrons in two-quantum dot cells as a function of geometry, potential and strength of interaction. For example, Fig.~\ref{fig:2dotsystem_evolution} shows an analog of occupancy oscillations from Fig.~\ref{fig:SEvsTB} for particle~1 in its respective double-quantum dot cell~$\alpha$ in the {\it presence} of electrostatic interaction with particle~2 in double-quantum dot~$\beta$ (refer to Fig.~\ref{fig:3dotsystem}(a) for the geometry and the arrangement of double-quantum dot cells). We begin with the configuration where particles~1 and~2 are both located in the left quantum dots of their respective double-quantum dot cells. Due to the repelling action of interaction, the occupancy oscillations are disturbed, and it is less likely to localize particle~1 in the right quantum dot of double-quantum dot $\alpha$. Figure~\ref{fig:2dotsystem_evolution} also shows the least and the most probable configurations of such a system.


\subsection{Three and More Interacting Qubits. Generating the wavefunctions and Matrices for an Arbitrary Number of Particles}

The generalization of the formalism, allowing an extension of the model to a multi-particle system, can be derived from these examples as follows. The wavefunction is expressed as a tensor product defined in the combined Hilbert space $\cal{H}^{(\otimes N)}$: 
\begin{equation}\label{eq:multi-particle-wavef}
\ket{\Psi} = \sum_{n_\alpha=1,0}\sum_{n_\beta=1,0}\ldots \sum_{n_\omega=1,0} c_{n_\alpha\,n_\beta...n_\omega}(t) \ket{n_\alpha^{(\alpha)}\,n_\beta^{(\beta)}\ldots n_\omega^{(\omega)}}
\end{equation}
and the corresponding Hamiltonian is constructed as follows:
\begin{equation}\label{eq:multi-particle-H}
\begin{split}
\hat {\text H} = & \hat {\text H}^{(\omega)}\otimes \hat {\text I}\otimes\ldots...\otimes \hat {\text I} + \ldots + \hat {\text I} \otimes \ldots \hat {\text I} \otimes \hat {\text H}^{(\alpha)}
\end{split}
\end{equation}

\begin{figure}[t!]
\includegraphics[width=\columnwidth]{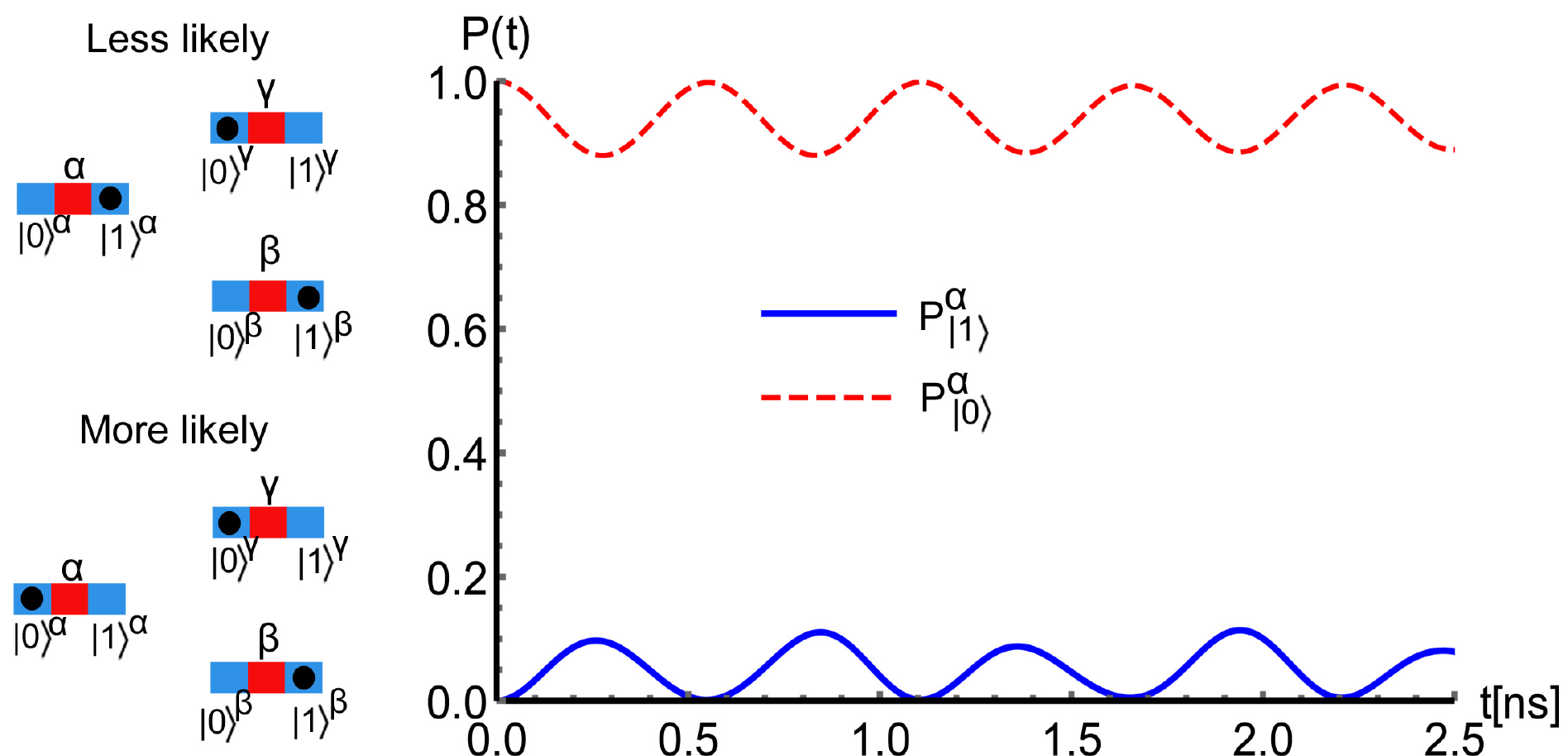}
\centering
\caption{Three interacting qubits: evolution of occupancy oscillations of particle~1 in double-quantum dot $\alpha$ expressed as probabilities $P^{\alpha}_{\ket{1}}$ and $P^{\alpha}_{\ket{0}}$ to locate the particle in the left or in the right quantum dot of the structure in the presence of interaction with particle~2 in double-quantum dot~$\beta$ and particle~3 in double-quantum dot $\gamma$. The scheme on the left shows the least and most probable configurations of the particles.}
\label{fig:3dotsystem_evolution}
\end{figure}

Expressions~\eqref{eq:multi-particle-wavef} and~\eqref{eq:multi-particle-H} allow us to assemble and solve the matrix equation~\eqref{eq:System_of_equations} for an arbitrary number of charge qubits of a given geometry. As an example, Fig.~\ref{fig:3dotsystem_evolution} visualizes a simulation for a three-electron system, where we plot an analog of occupancy oscillations for three interacting qubits. As expected, the occupancy oscillations are disturbed even more compared to Fig.~\ref{fig:2dotsystem_evolution} due to the presence of one additional particle and its repelling action. This approach can be easily automated and used to simulate interacting qubits and quantum gates in an environment compatible with circuit design and simulations.

\section{Von Neumann Entanglement Entropy in the Context of Semiconductor Charge Qubits}

To conclude this study, we shall also discuss whether it is possible to entangle two charge qubits when they interact electrostatically through the means of the Coulomb force. If entanglement is feasible for such qubits, this would mean that conventional quantum computing operations could be implemented on charge qubits. We recognize that there exist a number of methods to investigate entanglement. For example, the correlator as defined in Ref.~\cite{Emary_2009} is a direct derivation of the Bell inequality between two-qubits (DQDs). Both the Von Neumann entropy and this correlation function will result in the same conclusion of such a system in terms of simulation results. In~\cite{Pashkin_2003_Entanglement}, the concept of concurrence is defined for a pair of qubits.
As a measure of entanglement in this study, we will use the Von Neumann entropy. It is a bipartile quantity requiring to divide the original quantum system into two sub-parts. The aim of the analysis is to understand whether the sub-parts of the original system are in separable states or not.

\subsection{System of One Charge Qubit}

To understand how the entropy of entanglement works, let us start with a {\it single} qubit in a double-quantum dot system. For a bipartition of the single qubit system, we will use the following notation (this will help us in avoiding confusion with the notation introduced previously for the multi-electron case). We will denote the situation when the electron is in the left quantum dot of the double-quantum dot system as sub-part `$a$', and, likewise, for the right quantum dot as sub-part `$b$'. Subsequently, the state of one qubit can be conveniently written in the form similar to the multi-particle formalism we introduced before:
\begin{equation}
\ket{\psi} = c_{10} \ket{1^{a}0^{b}} + c_{01} \ket{0^{a}1^{b}}
\label{eq:qubit1}
\end{equation}
Here, $c_{10}(t)$ and $c_{01}(t)$ are the probability amplitudes in the localized position basis. This form should be interpreted as follows: $0^a$ means that there is no electron in sub-part $a$ of the system (i.e., in the left quantum dot) and $1^a$ means that the electron is present in sub-part $a$ of the system. The same applies when the index $b$ is used. For obvious reasons, the states $\ket {1^a 1^b}$, $\ket{0^a 0^b}$ or similar are not normally possible, as that would make the system physically corrupted.

To describe the Von Neumann entanglement entropy, we will use the density matrix formalism. In the position basis, noting equation~\eqref{eq:qubit1}, we write the density operator for the full system (no bipartition applied yet) in the following form:
\begin{equation}
\begin{split}
\hat{\rho}_{ab}&=\ket{\psi}\bra{\psi}=\\
&=
c_{10}c_{10}^{*} \ket{1^a 0^b} \bra{1^a 0^b} +
c_{01}c_{01}^{*} \ket{0^a 1^b} \bra{0^a 1^b}
\end{split}
\label{eq:xxx1}
\end{equation}
Equivalently, the density matrix operator can be written in a matrix representation, known as the density matrix $\boldsymbol{\rho}$, with elements $\rho_{ab}=[\boldsymbol{\rho}]_{ab}$ as follows,
\begin{equation}
\begin{matrix}
& &\hspace{11mm}\ket{1^a 0^b} & \ket{0^a 1^b}\end{matrix}\nonumber\\
\boldsymbol{\rho}
=\begin{matrix}\bra{1^a 0^b} \\ \bra{0^a 1^b} \end{matrix} \hspace{2mm} \begin{bmatrix} |c_{10}|^2 & 0 \\ 0 & |c_{01}|^2 \end{bmatrix}
\label{eq:xxx2}
\end{equation}

By dividing the system into two parts, $a$ and $b$, we can now introduce the reduced density operators $\hat{\rho}_{a}$ and $\hat{\rho}_{b}$ via the partial trace as follows:
\begin{equation}\label{eq:rdm_formula}
\begin{split}
\hat{\rho}_{a} = \bra{0^b} \hat{\rho}_{ab} \ket{0^b} + \bra{1^b} \hat{\rho}_{ab} \ket{1^b}\\
\hat{\rho}_{b} = \bra{0^a} \hat{\rho}_{ab} \ket{0^a} + \bra{1^a} \hat{\rho}_{ab} \ket{1^a}
\end{split}
\end{equation}
which describe the state of each sub-part, tracing out the complement. For example, in the matrix form, the reduced density matrix of sub-part $a$ takes the following form, which is equivalent to the reduced density matrix
\begin{equation}
\begin{matrix}
& &\hspace{9mm} \ket{0^a} && \ket{1^a}\end{matrix} \nonumber \\
\boldsymbol{\rho_{a}} =
\begin{matrix} \bra{0^a} \\ \bra{1^a} \end{matrix} \hspace{2mm}
\begin{bmatrix}
|c_{01}|^2& 0 \\ 0 & |c_{10}|^2 \end{bmatrix}
\label{eq:rdm}
\end{equation}
Then, the Von Neumann entanglement entropy $S_{\text{N}}$ is defined as follows:
\begin{equation}\label{eq:SNeumann}
S_{\text{N}} = - \text{Tr} (\hat{\rho}_a \ln\hat{\rho}_a) = - \text{Tr} (\hat{\rho}_b \ln\hat{\rho}_b)
\end{equation}

\begin{figure}[b!]
\includegraphics[width=.7\columnwidth]{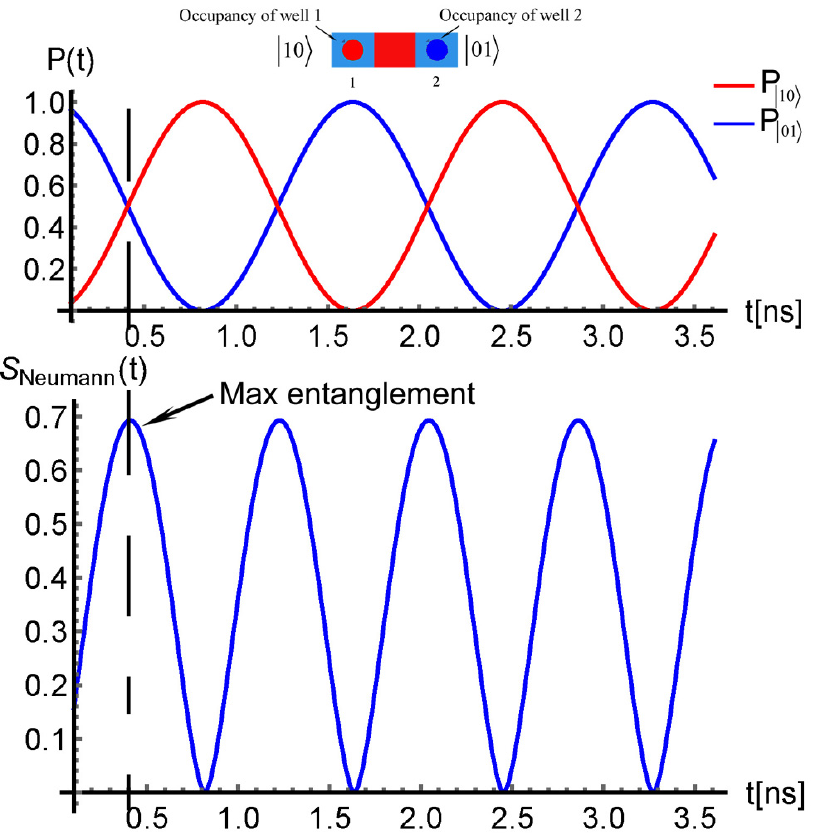}
\centering
\caption{Von Neumann entanglement entropy calculated for two states of a single charge qubit. Entanglement entropy is at maximum when the probability of occupying each of the states is the same. Vice versa, entanglement entropy is zero when the electron actualises in one of the quantum dots. 
}
\label{fig:SNeumann1qubit}
\end{figure}

While the entanglement entropy is not particularly useful for one electron or one qubit, it can provide some interesting insights to its meaning. As an illustration, the Von Neumann entropy applied to one qubit in a superposition state is visualized in Fig.~\ref{fig:SNeumann1qubit}. One can clearly see that the entropy is maximal, $S_{\text{N}}=\ln{2}$, when the qubit is found in one of the two states: $(\ket{1^a0^b} + \ket{0^a1^b})/\sqrt{2}$ or $(\ket{1^a0^b} - \ket{0^a1^b})/\sqrt{2}$. Recalling that we are operating in the localized basis, this superposition of the localized states corresponds to the situation when the electron is \emph{maximally delocalized}. We also note that in the localized basis a qubit displays occupancy oscillations although the period can vary by several orders-of-magnitude as a function of $U_B$. Hence, the Von Neumann entanglement entropy is generally a function of time.

\subsection{System of Two Interacting Qubits}

Having understood the Von Neumann entropy applied to the simplistic case of one qubit, we can now apply it to two interacting qubits. We will now return to our standard notation for double-quantum dots as used in Sec.~\ref{sec:two-qubits}. In this case, sub-part `$\alpha$' will denote an electron in the first double-quantum dot and sub-part `$\beta$' will denote an electron in the second double-quantum dot. The wavefunction is the same as given by formula~\eqref{eq:2wavefunc}:
\begin{equation}\label{eq:2wavefunc2}
\begin{split}
\ket{\Psi}=c_{11} \ket{1^\alpha 1^\beta} + c_{10} \ket{1^\alpha 0^\beta} + c_{01} \ket{0^\alpha 1^\beta} + c_{00} \ket{0^\alpha 0^{\beta}}
\end{split}
\end{equation}
The density operator becomes:
\begin{equation}
\begin{split}
\hat{\rho}=
&c_{11}c_{11}^{*} \ket{1^\alpha 1^\beta} \bra{1^\alpha 1^\beta} +
c_{10}c_{10}^{*} \ket{1^\alpha 0^\beta} \bra{1^\alpha 0^\beta} +\\
&c_{01}c_{01}^{*} \ket{0^\alpha 1^\beta} \bra{0^\alpha 1^\beta}
+ c_{00}c_{00}^{*} \ket{0^\alpha 0^{\beta}} \bra{0^\alpha 0^{\beta}}
\end{split}
\label{eq:rho_matrix}
\end{equation}
Equivalently, the density matrix can be written as follows:
\begin{equation}
\begin{matrix}
&\hspace{11mm}\ket{1^\alpha 1^\beta} & \hspace{-2mm} \ket{1^\alpha 0^\beta} & \hspace{-2mm} \ket{0^\alpha 1^\beta} & \hspace{-2mm} \ket{0^\alpha 0^{\beta}} \end{matrix} \nonumber \\
\boldsymbol{\rho} =
\begin{matrix}
\bra{1^\alpha 1^\beta}\\ \bra{1^\alpha 0^\beta} \\
\bra{0^\alpha 1^\beta} \\\bra{0^\alpha 0^\beta}
\end{matrix}
\begin{bmatrix} |c_{11}|^2 & 0 & 0 & 0 \\
0 & |c_{10}|^2 & 0 & 0 \\
0 & 0 & |c_{01}|^2 & 0\\
0 & 0 & 0 & |c_{00}|^2
\end{bmatrix}
\label{eq:xxx3}
\end{equation}
Applying a bipartition to the system, we obtain the reduced density matrices. For example, following Eq.~\eqref{eq:rdm_formula}, the reduced density matrix for sub-part $\alpha$ can be written as follows:
%
%
\begin{equation}
\begin{matrix}
 \hspace{12mm} \ket{0^\alpha} & & & & & \ket{1^\alpha}\end{matrix}\\
\boldsymbol{\rho_{\alpha}} =
\begin{matrix}
\bra{0^\alpha}\\
\bra{1^\alpha}
\end{matrix}
\begin{bmatrix} |c_{01}|^2 + |c_{00}|^2& 0 \\
0 & |c_{10}|^2 + |c_{11}|^2 \\
\end{bmatrix}
\label{eq:xxx5}
\end{equation}
We note here that the entries of the reduced density matrix can be easily understood. In the matrix above, the first non-zero entry gives the probability of finding sub-part $\alpha$ of the quantum system in state $\ket{0}$, while the second non-zero entry gives the probability of finding it in state $\ket{1}$, \emph{regardless} of the state of the second electron. These entries, as a matter of fact, were plotted in the graphs of Fig.~\ref{fig:2dotsystem_evolution}. As a final step, the Von Neumann entanglement entropy is calculated using the same formula~\eqref{eq:SNeumann}.

\begin{figure}[t!]
\includegraphics[width=.75\columnwidth]{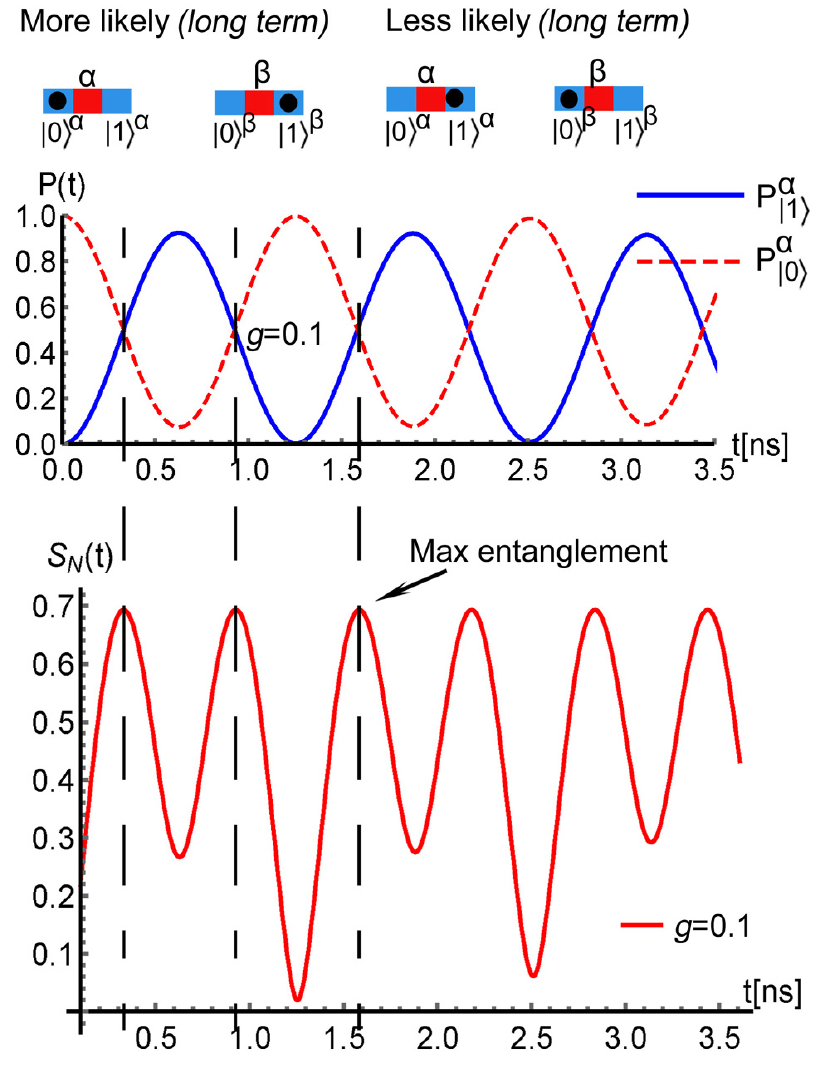}
\centering
\caption{Von Neumann entanglement entropy $S_{\text{N}}(t)$ of two electrostatically interacting qubits as a function of time calculated at electrostatic screening coefficient $g=0.1$.}
\label{fig:SNeumann2qubits_g01}
\end{figure}

\begin{figure}[t!]
\includegraphics[width=.70\columnwidth]{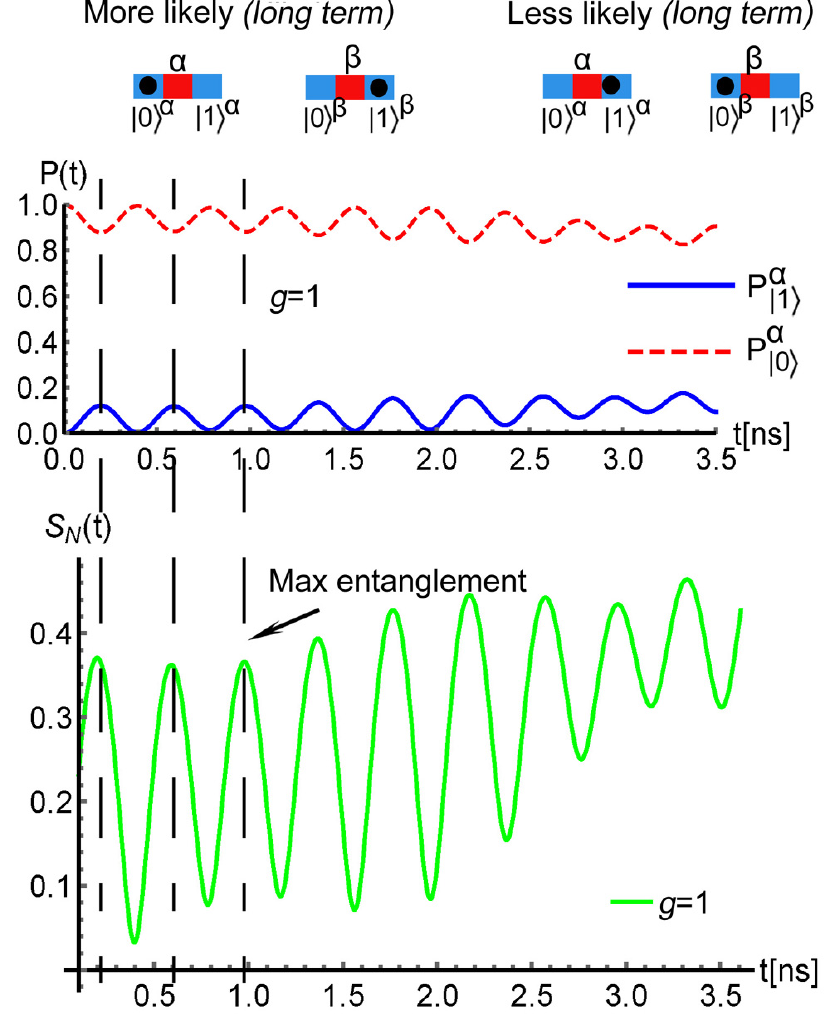}
\centering
\caption{Initial Von Neumann entanglement entropy $S_{\text{N}}(t)$ of two electrostatically interacting qubits as a function of time calculated at electrostatic screening coefficient $g=1$.}
\label{fig:SNeumann2qubits_g1}
\end{figure}

\begin{figure}[t!]
\includegraphics[width=.8\columnwidth]{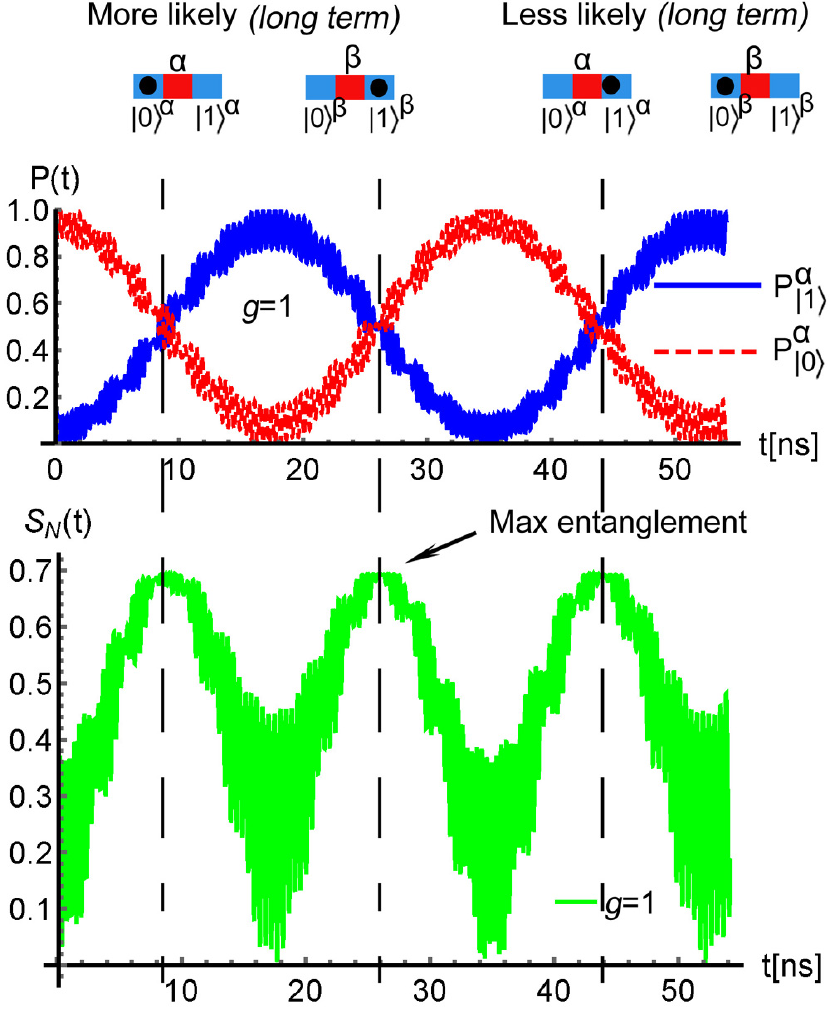}
\centering
\caption{Von Neumann entanglement entropy $S_{\text{N}}(t)$ of two electrostatically interacting qubits as a function of time over a longer time stretch calculated at electrostatic screening coefficient $g=1$.}
\label{fig:SNeumann2qubits_g1_long_time}
\end{figure}

\begin{figure}[t!]
\includegraphics[width=.8\columnwidth]{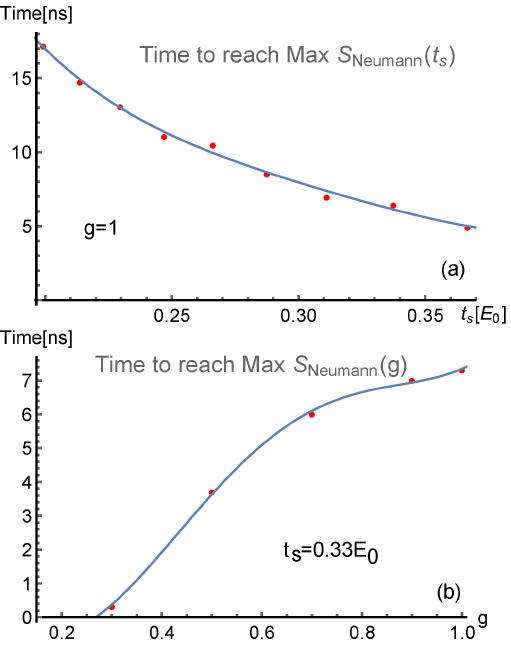}
\centering
\caption{(a) Time to reach maximum entanglement $S_{\text{N}}(t)$ as a function of the tunneling probability $t_h$ for a fixed value of the screening coefficient $g=1$. (b) Time to reach maximum entanglement $S_{\text{N}}(t)$ as a function of the screening coefficient $g$ for a fixed value of the tunneling probability $t_h=0.33 E_0$.}
\label{fig:Time_to_max_entanglement}
\end{figure}

Figure~\ref{fig:SNeumann2qubits_g01} shows the entanglement entropy $S_{\text{N}}(t)$ calculated for two qubits (each containing an electron) interacting electrostatically though the Coulomb's force. The Hamiltonian operator is given by expression~\eqref{eq:h-many-particles}, and we note that we use the effective screening constant $g$ to account for possible screening effects in solid-state structures~\cite{Kruchinin_2010}. Fig.~\ref{fig:SNeumann2qubits_g01} also shows the occupancy oscillations experienced by sub-part $\alpha$ (i.e., the electron in the first double-quantum dot) in the presence of the second electron when the electrostatic interaction between them is rather weak ($g = 0.1$). In this case, as was discussed earlier, the electron experiences large-amplitude occupancy oscillations, with $\sim$100\% probability to be eventually found in the left quantum dot and a bit less in the right quantum dot. Reciprocally, there are instances of time when the electron is delocalised, and the entanglement entropy reaches its maximum. For comparison, Fig.~\ref{fig:SNeumann2qubits_g1} shows an example of strong interaction between the two qubits ($g = 1$), whilst Fig.~\ref{fig:SNeumann2qubits_g1_long_time} shows the same example for a longer time stretch. The occupancy oscillations are disturbed, and the electron $\alpha$ is initially found mostly in the left quantum dot due to the strong electrostatic repulsion. As a consequence, entanglement entropy decreases. However, the dynamics follow the same pattern (with a different frequency), when one visualizes the same plot for a longer time period. We conclude that if the interaction between the two qubits is strong, it leads to the localisation of electrons and hence it reduces their entanglement.

To conclude this section, the time to reach maximum entanglement between the two qubits is plotted in Fig.~\ref{fig:Time_to_max_entanglement}(a) as a function of the tunneling probability $t_h$ and for a fixed value of the screening coefficient $g=1$. It is evident that the higher the hopping term the shorter the time needed. Lastly, we also plot in Fig.~\ref{fig:Time_to_max_entanglement}(b) the time to reach maximum entanglement between the two qubits as a function of the screening coefficient $g$, for a fixed value of the tunneling probability $t_h=0.33 E_0$. The higher the screening coefficient the longer it takes for the system to reach the maximum entanglement.

\subsection{System of Two Interacting Registers}

\begin{figure}[t!]
\includegraphics[width=.8\columnwidth]{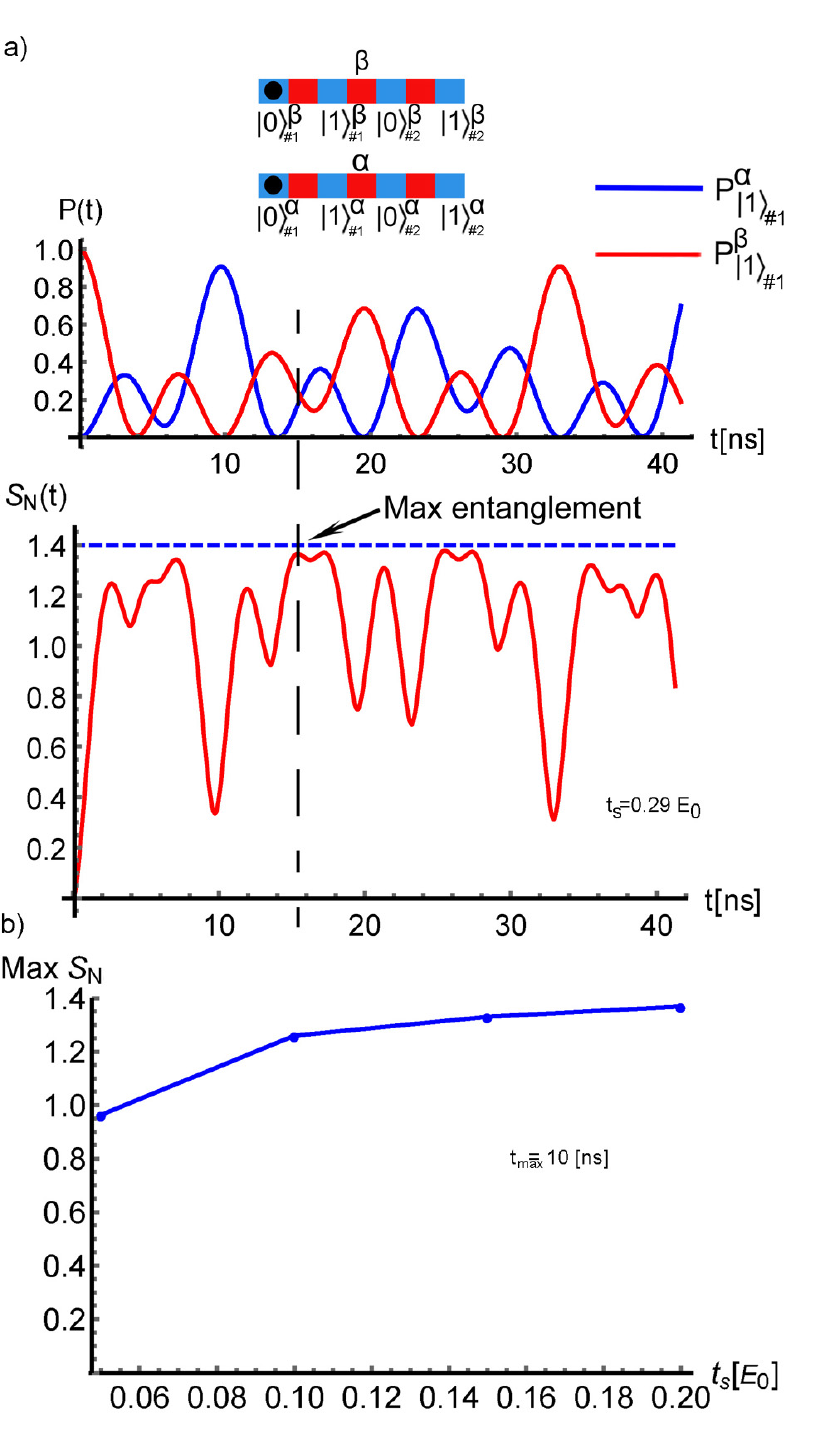}
\centering
\caption{(a) Von Neumann entanglement entropy $S_{\text{N}}(t)$ between two single-electron registers interacting electrostatically via Coulomb interaction. Each line consists of two DQDs which corresponds to two qubits, denoted as qubit\#1 and qubit\#2; (b) Time to reach maximum entanglement $S_{\text{N}}(t)$ as a function of the tunneling probability $t_h$ in a given time duration, $t_{max}=10$~ns.}
\label{fig:SN_4_dots}
\end{figure}

In this section, we will examine a system of two coupled registers each consisting of four QDs, as depicted in Fig.~\ref{fig:SN_4_dots}(a). We will assume one particle in each line interacting electrostatically via the Coulomb force with one particle on the other line. In this case, sub-part `$\alpha$' denotes the first line and sub-part `$\beta$' denotes the second line. Each line includes four dots (or two DQDs) and can be seen as a qudit. The wavefunction of the system is:
\begin{equation}
\begin{split}
\ket{\Psi}= \sum_{n_A=0_{\#1},1_{\#1},0_{\#2},1_{\#2} }\:\:\sum_{n_B=0_{\#1},1_{\#1},0_{\#2},1_{\#2}} c_{n_A n_B}\ket{n_A^{(A)}n_B^{(B)}}
\end{split}
\label{eq:coupled_lines}
\end{equation}
where we again assume four quantum states for each particle.

The reduced density matrix for sub-part $\alpha$ can be written as follows:
\begin{equation}
\hat{\rho}_{\alpha} = \bra{0_{\#1}^{\beta}} \hat{\rho} \ket{0_{\#1}^{\beta}} + \bra{1_{\#1}^{\beta}} \hat{\rho} \ket{1_{\#1}^{\beta}} + \bra{0_{\#2}^{\beta}} \hat{\rho} \ket{0_{\#2}^{\beta}}+ \bra{1_{\#2}^{\beta}} \hat{\rho} \ket{1_{\#2}^{\beta}}
\end{equation}
Let us now assume the maximally entangled Bell state~\cite{Terhal_2000, Giounanlis_2019mdpi}:
\begin{equation}
\ket{\Phi} = \frac{1}{\sqrt{2}}\left(\ket{0^{\alpha}0^{\beta}} + \ket{1^{\alpha}1^{\beta}}\right)
\end{equation}
which is defined generally between any two systems, $\alpha$ and $\beta$.
Writing this state in the basis of our system, we get
\begin{align}
\ket{\Phi} = \frac{1}{\sqrt{2}}\Bigl(\ket{0_{\#1}^{\alpha}0_{\#1}^{\beta}}+\ket{0_{\#2}^{\alpha}0_{\#2}^{\beta}} + \ket{0_{\#1}^{\alpha}0_{\#2}^{\beta}} + \ket{0_{\#2}^{\alpha}0_{\#1}^{\beta}}+\nonumber\\
+\ket{1_{\#1}^{\alpha}1_{\#1}^{\beta}}+\ket{1_{\#2}^{\alpha}1_{\#2}^{\beta}} + \ket{1_{\#1}^{\alpha}1_{\#2}^{\beta}} + \ket{1_{\#2}^{\alpha}1_{\#1}^{\beta}}\Bigr)
\end{align}
and
\begin{equation}
\hat{\rho}_{\alpha} = \text{tr}_{\beta}\left(\bra{\Phi}\ket{\Phi}\right) = \frac{1}{2} {\textbf I_4}
\end{equation}
From~\eqref{eq:SNeumann} we can calculate $S_N=2 \ln{2}$.

In Fig.~\ref{fig:SN_4_dots}(a), the Von Neumann entanglement entropy $S_{\text{N}}(t)$ is plotted as defined between the two single-electron registers interacting electrostatically via the Coulomb interaction. Each line consists of two DQs which correspond to two qubits, denoted as qubit \#1 and qubit \#2. Interestingly, it is visible that an almost maximally entangled state can be achieved in this case ($\sim 2\ln 2$) for the selected parameters. In principle, the maximum entanglement is harder to achieve as the spatial degrees of freedom for each particle increases. Finally, the maximum entanglement achieved in the system $S_{\text{N}}$ as a function of the tunneling probability $t_h$, in a given time duration $t_{\text{max}}=10$[ns], is plotted in Fig.~\ref{fig:SN_4_dots}(b). As the tunneling probability $t_h$ increases, the system can get maximally entangled.

\section{Decoherence and Fidelity of Charge Qubits}
\label{Sec:decoherence}

Qubits based on nuclear spin, electron spin or combination of charge-spin (hybrid) appear to be the best candidates for quantum computing from the point of view of achievable decoherence times~\cite{Weichselbaum_2004, Giounanlis_2019}.
As reported in the literature, charge-based qubits have much shorter decoherence times (within a range of 50~ns to 1~$\mu$s). However, reasonably high {\it practical} qubit fidelity can still be achieved by using very fast state flip times as explained below. The 22-nm FDSOI CMOS process used to develop the quantum structures considered in this study has transition frequencies ($f_T$) in the several hundreds of gigahertz range~\cite{Bonen_2018}. Even with a 50~ns decoherence time, charge-based qubits could still perform over a thousand operations per useful coherence duration. This is an advantage that can compensate for the charge qubits' shorter decoherence time. The FDSOI process has a roadmap for a 12-nm feature node that should provide state flip times as low as 10~ps and hence can facilitate high quality quantum operation. We also note that the measurement/readout of {\it spin} qubits is very challenging due to their need to operate with narrowband microwave electromagnetic fields which are inherently slow (tens of ns-level access time) and consume rather high power (over 10\,mW per qubit)~\cite{Leipold_2019, Leipold_2019b, Voinigescu_2019}. Consequently, being able to operate on charge qubits 2--3 orders-of-magnitude faster can compensate for their 2--3 orders-of-magnitude decoherence time handicap.

Lastly, we should also mention that by choosing the state and the physical parameters properly, the dephasing can be largely suppressed in a quantum-dot array~\cite{Yang_2019, Petersson_2010,Chan_2019, Ghosh_2017b}. Therefore, we can conclude that the relatively short decoherence time is not expected to prohibit the application of charge qubits for quantum computing.

\section{Conclusions}

This paper provides a formal definition, robustness analysis and discussion on the control of a charge qubit intended for semiconductor implementation in scalable CMOS quantum computers. The construction of the charge qubit requires maximally localized functions, and we show such functions for double-quantum dot structures with dimensions corresponding to a 22-nm FDSOI CMOS technology. We also discuss how an individual qubit can be manipulated in terms of the two angles of the Bloch sphere.

Based on the electrostatic nature of the qubit, we demonstrate how to build a tight-binding model of one and multiple interacting qubits from first principles of the Schr\"odinger formalism. We provide all required formulae to calculate the maximally localized functions and entries of the Hamiltonian matrix in the presence of interaction between qubits. We use four illustrative examples to demonstrate interaction of electrons in three cases of interest and discuss how to build a model for many-electron (qubit) system. Finally, we use the Von Neumann entanglement entropy in the context of charge qubits to show that the electrostatically interacting electrons in these qubits can be entangled.

\section{Acknowledgment}
Elena Blokhina and Panagiotis Giounanlis contributed equally to this work.
\balance

\bibliographystyle{IEEEtran}
\bibliography{bib_rev2_clean}

\EOD
\end{document}